\documentclass[landscape,fleqn]{article}
\usepackage{latexsym}
\usepackage{graphicx}
\usepackage{amsmath}
\usepackage{amsfonts}
\usepackage{amssymb}%

\begin{document}
\title{Fourth-order perturbative extension of the single-double excitation
coupled-cluster method,\\
Part II:  Angular reduction}
\author{Andrei Derevianko\\
Physics Department, University of Nevada, Reno, NV 89557 }
\maketitle

\begin{abstract}
We tabulate angularly reduced fourth-order many-body corrections
to matrix elements for univalent atoms, derived in [A. Derevianko and E.D. Emmons,
Phys. Rev. A
{\bf 65 }, 052115 (2002)].
In particular we focus on practically important
diagrams complementary to those included in the coupled cluster method
truncated at single and double excitations.
Derivation  and angular reduction of a large number of diagrams
have been carried out with the help of symbolic algebra software.
\end{abstract}

\section{Generalities}

This e-print serves as an electronic supplement to Ref.~\cite{DerEmm02}.
In that
paper we derived fourth-order many-body corrections to matrix elements for univalent atoms.
Based on the derived diagrams we proposed next-generation many-body method for
calculating atomic properties such as parity-violating amplitudes.
Here I carry out the next necessary
step required in a practical implementation of this method --- angular reduction of the relevant
diagrams.

\begin{figure}[h]
\begin{center}
\includegraphics*[scale=0.65]{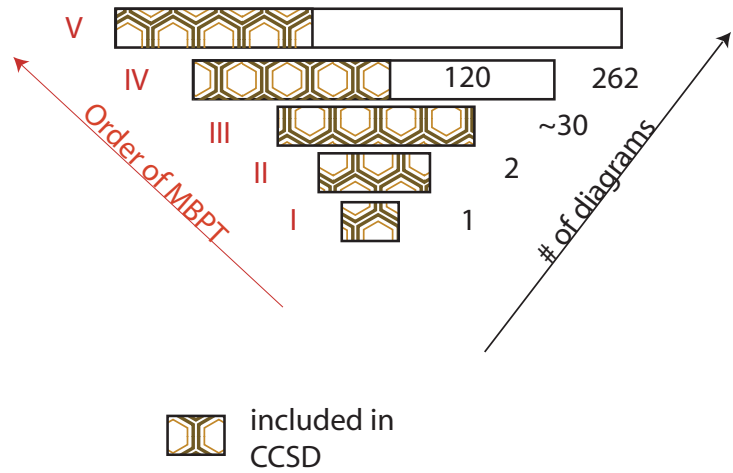}
\caption{ \small CCSD vs order-by-order MBPT for one-particle matrix elements. \label{FigCCSDvsMBPT} }
\end{center}
\end{figure}

In Ref.~\cite{DerEmm02} the fourth-order diagrams were
classified using coupled-cluster-inspired separation
into contributions from $n$-particle excitations from the lowest-order
wavefunction.
It was found that the complete set of fourth-order diagrams involves only
connected single, double, and triple excitations and disconnected
quadruple excitations.
Approximately half of the fourth-order
diagrams is {\em not} accounted for by the popular coupled-cluster method~\cite{LinMor86} truncated
at single and double excitations (CCSD). To devise a practical scheme
capable of improving accuracies of the existing many-body methods,
we proposed to combine direct order-by-order
many-body perturbation theory (MBPT) with the truncated CCSD method.
This idea is illustrated in Fig.~\ref{FigCCSDvsMBPT}: the CCSD method recovers
all many-body diagrams up to the third order of MBPT for matrix elements, but
misses contributions starting from the fourth order.
Such a fusion
of (truncated) all-order methods with order-by-order MBPT promises improved
accuracy in calculation of parity-violating effects for several practically interesting
atoms such as Cs, Fr, and with some modifications to Tl. It is worth noting that based on the
derived fourth-order diagrams
we also devised a partial summation scheme to  all orders of MBPT~\cite{DerEmm02}. The discussion
of that approach is beyond the scope of the present e-print.

We considered a matrix element $M_{wv}$ of non-scalar operator $Z$ between two states of
valence electron $w$ and $v$. The set of MBPT diagrams complementary to CCSD is entirely due to omitted
triple excitations from the reference Hartree-Fock determinant. We separated
these additional contributions into three major classes by noting
that triples enter the fourth order matrix element $\mathcal{M}^{(4)}_{wv}$  via
\begin{enumerate}
\item{ an {\em indirect} effect of triples on single and double excitations in the
third-order wavefunction --- we denote this class as $Z_{0\times 3}$,  }
\item{ {\em direct} contribution of triples to matrix elements
       ---  class $Z_{1\times 2}$,  }
\item{ correction to normalization --  $Z_\mathrm{norm}$.}
\end{enumerate}
Further these classes are broken into subclasses based on the nature of triples, so that
\begin{eqnarray}
\lefteqn{ \left( \mathcal{M}^{(4)}_{wv} \right)_\mathrm{non-CCSD} =
Z_{1 \times 2}(T_v) + Z_{1 \times 2}(T_c) + } \\
& &Z_{0 \times 3}(S_v[T_v]) + Z_{0 \times 3}(D_v[T_v]) +
 Z_{0 \times 3}(S_c[T_c]) + Z_{0 \times 3}(D_v[T_c]) + \nonumber \\
& & Z_\mathrm{norm}(T_v) \, . \nonumber
\end{eqnarray}
Here we distinguished between valence ($T_v$) and core ($T_c$) triples and
introduced a similar notation for singles ($S$) and doubles ($D$).
Notation like $S_v[T_c]$ stands for an effect of second-order core triples ($T_c$) on
third-order valence singles $S_v$.
The reader is referred to Ref.~\cite{DerEmm02} for further details and discussion.
Representative diagrams are shown in Fig.~\ref{FigZ4triples} and
algebraic expressions are tabulated in the Appendix of Ref.~\cite{DerEmm02}.

\begin{figure}[h]
\begin{center}
\includegraphics*[scale=0.65]{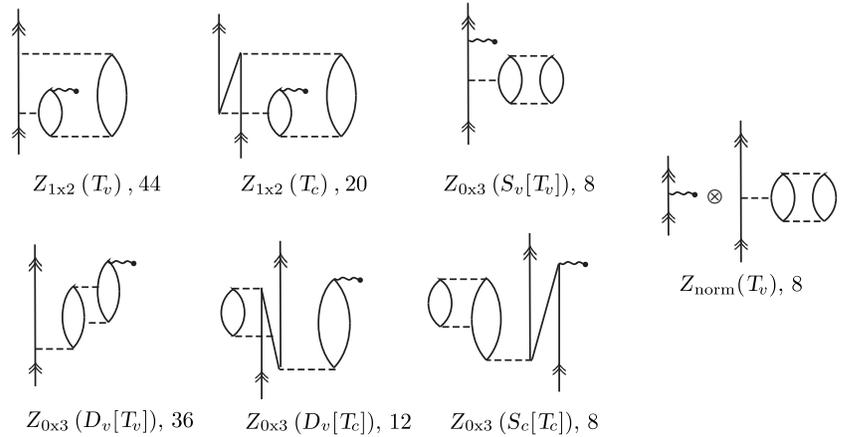}
\caption{ \small \label{FigZ4triples} Sample fourth-order diagrams involving triple
excitations. The one-particle
matrix element is denoted by a wavy horizontal line.
The number of contributions
for each class of diagrams is also shown;
direct, all possible exchange, and the conjugated graphs of a given
diagram were counted as a single contribution.}
\end{center}
\end{figure}

\subsection{Sample contribution and notation}
Here is a sample fourth-order  term  derived in Ref.~\cite{DerEmm02}
\begin{equation}
 Z_{0 \times 3}(S_c[T_c])=
-\sum_{abcmnr} \frac{{z_{bv}}{\tilde{g}}_{acnr}{g_{nrcm}}{\tilde{g}}_{mwab}}
     {({{\varepsilon }_w}-{{\varepsilon }_{b}})\,
       ({{\varepsilon }_{mw}}-{{\varepsilon }_{ab}})\,
       ({{\varepsilon }_{nrw}}-{{\varepsilon }_{abc}})}\, + {\rm 7\, additional \, terms} + \mathrm{h.c.s.}
       \label{Eqn_sample}
\end{equation}
In energy denominators, abbreviation
$\varepsilon_{xy\ldots z}$  stands for
$\varepsilon_{x} + \varepsilon_{y} + \cdots \varepsilon_{z}$, with $\varepsilon_{x}$
being single-particle Dirac-Hartree-Fock (DHF) energy.
Further, $g_{ijlk}$ are matrix elements of Coulomb interaction in the basis
of DHF orbitals $u_i(\mathbf{r})$
\begin{equation}
g_{ijkl}=\int u_{i}^{\dagger}\left(  \mathbf{r}\right)  u_{j}^{\dagger}\left(
\mathbf{r}^{\prime}\right)  \frac{1}{\left|  \mathbf{r}-\mathbf{r}^{\prime
}\right|  }u_{k}\left(  \mathbf{r}\right)  u_{l}\left(  \mathbf{r}^{\prime
}\right)  d^{3}r\,d^{3}r^{\prime} \, .
\label{Eqn_CoulMel}
\end{equation}
The quantities $\tilde{g}_{ijlk}$
are antisymmetric combinations $\tilde{g}_{ijlk}={g}_{ijlk}- {g}_{ijkl}$.
The summation is over single-particle DHF states,
labelled according to the following convention:
core orbitals are enumerated by letters $a,b,c,d$,  complementary
excited states are labelled by $m,n,r,s$,
and valence orbitals are denoted by $v$ and $w$.
Finally matrix elements of operator $\hat{Z}$ in the DHF basis are denoted $z_{ij}$
and the $\mathrm{h.c.s.}$ contribution is to be calculated by taking the hermitian conjugate
of all preceding terms and swapping labels $v$ and $w$.

\subsection{Angular reduction}
Having introduced building blocks of a many-body contribution to matrix elements,
now we proceed to angular reduction, which means carrying out a summation
over magnetic quantum numbers in a closed form.

One-particle DHF orbital may be conventionally represented as
\begin{equation}
u(\mathbf{r})=\frac{1}{r}\left(
\begin{array}
[c]{c}%
iP_{n\kappa}(r)\ \Omega_{\kappa m}(\hat{r})\\
Q_{n\kappa}(r)\ \Omega_{-\kappa m}(\hat{r})
\end{array}
\right) \, ,
\label{Eqn_bispinor}
\end{equation}
where $P$ and $Q$ are large and small radial components, $\kappa=(l-j)\left(
2j+1\right)$, and $\Omega_{\kappa m}$ is the spherical spinor.
Then in Eq.~(\ref{Eqn_sample}) a summation over an orbital $i$ encapsulates
summation over principal quantum number $n_i$, angular momentum $j_i$ (or $\kappa_i$),
and magnetic quantum numbers $m_i$.

The Wigner-Eckart (WE) theorem~\cite{Edm85} allows to ``peel off'' $m$-dependence of various matrix
elements. WE theorem states that
if an operator $Z^{(K)}_Q$ is the $Q^\mathrm{th}$  component of an irreducible tensor operator of rank $K$,
then the matrix element $\langle n_1 j_1 m_1| Z^{(K)}_M|  n_2 j_2 m_2 \rangle$ may be expressed as
\begin{equation}
\langle n_1 j_{1} m_{1}|Z^{(K)}_Q| n_2 j_{2} m_{2}\rangle=
\left(  -1\right)^{j_{1}-m_{1}}
\left(
\begin{array}
[c]{ccc}%
j_{1} & K & j_{2}\\
-m_{1} & Q & m_{2}%
\end{array}
\right)  \langle n_1 j_{1}||Z^{(K)}||n_2 j_{2}\rangle \, ,
\label{EqnWE}
\end{equation}
where $\langle n j | Z^{(K)}|  n' j' \rangle$ is a reduced matrix element.
Using the WE theorem and expansion of $1/|\mathbf{r} - \mathbf{r}' |$ into Legendre polynomials,
the Coulomb matrix element~(\ref{Eqn_CoulMel}) is
traditionally represented as
\begin{equation}
g_{abcd}=\sum_{LM}\left(  -1\right)  ^{L-M}\left(  -1\right)  ^{j_{a}-m_{a}%
}\left(
\begin{array}
[c]{ccc}%
j_{a} & L & j_{c}\\
-m_{a} & M & m_{c}%
\end{array}
\right)  \left(  -1\right)  ^{j_{b}-m_{b}}\left(
\begin{array}
[c]{ccc}%
j_{b} & L & j_{d}\\
-m_{b} & -M & m_{d}%
\end{array}
\right)  X_{L}(abcd) \, ,
\label{Eqn_greduce}
\end{equation}
where Coulomb integral
\[
X_{L}(abcd)=(-1)^{L}\langle\kappa_{a}||C^{\left(  L\right)  }||\kappa
_{c}\rangle\langle\kappa_{b}||C^{\left(  L\right)  }||\kappa_{d}\rangle
R_{L}(abcd)
\]
is defined in terms of reduced matrix element of normalized spherical harmonics $C^{(L)}$~\cite{VarMosKhe88}
and a Slater integral expressed in terms of radial components of single-particle orbitals
\[
R_{L}(abcd)=\int_{0}^{\infty}dr_{1}[P_{a}(r_{1})P_{c}(r_{1})+Q_{a}(r_{1}%
)Q_{c}(r_{1})]\int_{0}^{\infty}dr_{2}\frac{r_{<}^{L}}{r_{>}^{L+1}}[P_{b}%
(r_{2})P_{d}(r_{2})+Q_{b}(r_{2})Q_{d}(r_{2})
\]
with $r_{<} = \min(r_1, r_2)$ and $r_{>} = \max(r_1, r_2)$.
The anti-symmetrized combinations $\tilde{g}_{abcd}=g_{abcd}-g_{abdc}$ are
reduced similar to $g_{abcd}$ , except in Eq.(\ref{Eqn_greduce}) $X_{L}(abcd)$ is replaced with
\[
Z_{L}(abcd)=X_{L}(abcd)+[L]\sum_{L^{\prime}}\left\{
\begin{array}
[c]{ccc}%
b & d & k\\
a & c & k^{\prime}%
\end{array}
\right\}  X_{L^{\prime}}(bacd) \, .
\]
Here $[L]=2L+1$.
It is worth emphasizing that both $Z_{L}(abcd)$ and $X_{L}(abcd)$ do not
depend on magnetic quantum numbers of single-particle states.

Angular reduction, i.e. summation over magnetic quantum numbers of atomic single-particle orbitals
in many-body diagrams $\mathcal{M}_{wv}$ such as Eq.(\ref{Eqn_sample}) ,
leads to  many-body correction to {\em reduced} matrix elements, $\bar{M}_{wv}$, as prescribed by the WE theorem (\ref{EqnWE})
\begin{equation}
\mathcal{M}_{wv}=
\left(  -1\right)^{j_{w}-m_{w}}
\left(
\begin{array}
[c]{ccc}%
j_{w} & K & j_{v}\\
-m_{w} & Q & m_{v}%
\end{array}
\right) \overline{M}_{wv} \, ,
\end{equation}
where $K$ and  $Q$ are the rank and component of the underlying one-particle operator $Z$.
In symbolic calculations it is more convenient to invert this relation and compute
\[
 \overline{M}_{wv} = \sum_{m_w m_v Q} \left(  -1\right)^{j_{w}-m_{w}}
\left(
\begin{array}
[c]{ccc}%
j_{w} & K & j_{v}\\
-m_{w} & Q & m_{v}%
\end{array}
\right) \mathcal{M}_{wv} \, .
\]

To derive  many-body diagrams and carry out angular reduction
we developed a symbolic tool based on Mathematica~\cite{Wol99} and publicly
available angular reduction routine~\cite{Tak92}. This package
allows to work with MBPT expressions in an interactive regime.
For example, all the \LaTeX \  formulae tabulated in the Section~\ref{SecFormulae}
have been generated automatically.
Without the help of symbolic tools,
the sheer number of diagrams  in the fourth order of MBPT
would have made  the traditional ``pencil-and-paper'' approach unmanageable and error-prone.
The correctness of the developed code has been verified by repeating results of
angular reduction for the third-order corrections to matrix elements, tabulated by Johnson, Liu,
and Sapirstein~\cite{JohLiuSap96}.

The results of the angular reduction is given in Section~\ref{SecFormulae}.
In addition to the Coulomb matrix elements $X_L(abcd)$ and $X_L(abcd)$ already introduced,
we used the following notation.
Reduced matrix elements of a non-scalar one-particle operator $Z$ are denoted as
$\langle i || z || j \rangle$, $K$ is the rank of the operator, $(-1)^{a+\ldots} = (-1)^{j_a+\ldots} $,
$\delta _{\kappa }(a,b) = \delta_{\kappa_a, \kappa_b}$, and $[a] = 2\,j_a+1$.

As to the angular reduction of  $\mathrm{h.c.s.}$  terms, it is given simply by
adding a phase factor and swapping labels $w$ and $v$ in the main contribution $M_{wv}$
\begin{equation}
\overline{ \mathrm {h.c.s.} \, \left( M_{wv} \right) } = (-1)^{w-v}
\bar{M}_{wv} (w \leftrightarrow v) ,
\label{EqnAngRedhcs}
\end{equation}
provided reduced matrix element of one-particle operator satisfies
\begin{equation}
\langle a||z||b \rangle = (-1)^{a-b} \langle b ||z|| a  \rangle \, .
\label{EqRmelRequirement}
\end{equation}
The relation~(\ref{EqnAngRedhcs}) allows us to carry out angular reduction and code
only  half of the diagrams, which is of a great utility
considering a couple of hundreds of diagrams in the fourth order MBPT.
The requirement (\ref{EqRmelRequirement}) is not restrictive, it holds
for all practically important matrix elements: non-retarded electric and magnetic
multipoles, hyperfine, and parity-violating matrix elements.

To reiterate, in this e-print we have tabulated angularly reduced fourth-order corrections
to matrix elements for univalent atoms. Due to overwhelmingly large number of diagrams
we  focused on the diagrams complementary to those included in the coupled cluster method
truncated at single and double excitations. The derivation  of the diagrams and angular reduction
has been carried out with the help of symbolic algebra software. In the future we plan to extend
the suite to
automatically generate Fortran code for these contributions and to perform numerical evaluations.

I would like to thank W.R. Johnson, W.F. Perger, and K. Takada  for discussions.
This work has been supported in part by the National Science Foundation.

\bibliographystyle{alpha}
\bibliography{mypub,general,exact}

\section{Formulae}
\label{SecFormulae}


\subsection{$Z_{1\times 2}\left( T_c \right)$}
\[
Z_{1\times 2}\left( T_c \right) =
\]

\[
-\sum_{abcmnr}\sum_{L_{1}L_{2}}
    \frac{{\left( -1 \right) }^{c + L_{2} - r - v - w}\,
       \left\{ \begin{array}[c]{ccc}K & v &w  \\  b & L_{2} & L_{1}
        \end{array} \right\}\,\left\{ \begin{array}[c]{ccc}L_{1} & K &
        L_{2}  \\  a & m & n\end{array} \right\}\,Z_{L_{1}}(cbrv)\,
       Z_{L_{1}}(rncm)\,Z_{L_{2}}(mwab)\,\langle a||z||n\rangle}{
       ({{\varepsilon }_{mw}}-{{\varepsilon }_{ab}})\,
       ({{\varepsilon }_{rv}}-{{\varepsilon }_{bc}})\,
       ({{\varepsilon }_{nrw}}-{{\varepsilon }_{abc}})\,[L_{1}]}\,+
\]
\[
-\left( \frac{1}{[K]} \right) \sum_{abcmnr}\sum_{L_{1}}
    \frac{{\left( -1 \right) }^{a + b + c + L_{1} - m - n - r}\,
       \left\{ \begin{array}[c]{ccc}w & L_{1} &m  \\  b & K & v
        \end{array} \right\}\,Z_{K}(mnba)\,Z_{L_{1}}(cbrv)\,Z_{L_{1}}(rwcm)\,
       \langle a||z||n\rangle}{({{\varepsilon }_{mn}}-{{\varepsilon }_{ab}}
        )\,({{\varepsilon }_{rv}}-{{\varepsilon }_{bc}})\,
       ({{\varepsilon }_{nrw}}-{{\varepsilon }_{abc}})\,[L_{1}]}\,+
\]
\[
-\sum_{abcmnr}\sum_{L_{1}L_{3}}
    \frac{{\left( -1 \right) }^{b + K + L_{1} - n - v - w}\,
       \left\{ \begin{array}[c]{ccc}K & w &v  \\  c & L_{1} & L_{3}
        \end{array} \right\}\,\left\{ \begin{array}[c]{ccc}L_{3} & K &
        L_{1}  \\  a & m & r\end{array} \right\}\,Z_{L_{1}}(bcnv)\,
       Z_{L_{1}}(mnab)\,Z_{L_{3}}(wrcm)\,\langle a||z||r\rangle}{
       ({{\varepsilon }_{mn}}-{{\varepsilon }_{ab}})\,
       ({{\varepsilon }_{nv}}-{{\varepsilon }_{bc}})\,
       ({{\varepsilon }_{nrw}}-{{\varepsilon }_{abc}})\,[L_{1}]}\,+
\]
\[
\frac{1}{[w]}\sum_{abcmnr}\sum_{L_{1}L_{3}}
    \frac{\delta _{\kappa }(a,w)\,\delta _{\kappa }(b,m)\,
       {\left( -1 \right) }^{c + L_{3} - n + r + w}\,X_{L_{1}}(nrcm)\,
       Z_{L_{1}}(cbnr)\,Z_{L_{3}}(mwab)\,\langle a||z||v\rangle}{
       ({{\varepsilon }_{mw}}-{{\varepsilon }_{ab}})\,
       ({{\varepsilon }_{nr}}-{{\varepsilon }_{bc}})\,
       ({{\varepsilon }_{nrw}}-{{\varepsilon }_{abc}})\,[b]\,[L_{1}]}\,+
\]
\[
-\left( \frac{1}{[w]} \right) \sum_{abcmnr}\sum_{L_{1}L_{2}}
    \frac{\delta _{\kappa }(a,w)\,
       {\left( -1 \right) }^{b + c + L_{2} + m - n - r + w}\,
       Z_{L_{1}}(bcnr)\,Z_{L_{1}}(mnab)\,Z_{L_{2}}(rwcm)\,
       \langle a||z||v\rangle}{({{\varepsilon }_{mn}}-{{\varepsilon }_{ab}}
        )\,({{\varepsilon }_{nr}}-{{\varepsilon }_{bc}})\,
       ({{\varepsilon }_{nrw}}-{{\varepsilon }_{abc}})\,{[L_{2}]}^2}\,+
\]
\[
\frac{1}{[K]}\sum_{abcdmn}\sum_{L_{1}L_{2}}
    \frac{{\left( -1 \right) }^{b - m - v - w}\,
       \left\{ \begin{array}[c]{ccc}K & v &w  \\  d & L_{1} & L_{2}
        \end{array} \right\}\,\left\{ \begin{array}[c]{ccc}L_{2} & K &
        L_{1}  \\  a & c & n\end{array} \right\}\,X_{L_{1}}(awcd)\,
       Z_{K}(mnba)\,Z_{L_{2}}(cdnv)\,\langle b||z||m\rangle}{(
        {{\varepsilon }_{mn}}-{{\varepsilon }_{ab}})\,
       ({{\varepsilon }_{nv}}-{{\varepsilon }_{cd}})\,
       ({{\varepsilon }_{mnw}}-{{\varepsilon }_{bcd}})}\,+
\]
\[
\frac{1}{[K]\,[v]}\sum_{abcdmn}\sum_{L_{1}}
    \frac{\delta _{\kappa }(a,v)\,
       {\left( -1 \right) }^{b - c + d + K - m - n - v}\,X_{L_{1}}(ancd)\,
       Z_{K}(mwba)\,Z_{L_{1}}(dcnv)\,\langle b||z||m\rangle}{(
        {{\varepsilon }_{mw}}-{{\varepsilon }_{ab}})\,
       ({{\varepsilon }_{nv}}-{{\varepsilon }_{cd}})\,
       ({{\varepsilon }_{mnw}}-{{\varepsilon }_{bcd}})\,[L_{1}]}\,+
\]
\[
\sum_{abcdmn}\sum_{L_{1}L_{2}L_{3}}
   \frac{{\left( -1 \right) }^{-b + K + L_{3} - v}\,
      \left\{ \begin{array}[c]{ccc}L_{2} & L_{3} &L_{1}  \\  a & c & m
       \end{array} \right\}\,\left\{ \begin{array}[c]{ccc}L_{3} & v &n
         \\  d & L_{1} & L_{2}\end{array} \right\}\,
      \left\{ \begin{array}[c]{ccc}v & L_{3} &n  \\  b & K & w
       \end{array} \right\}\,X_{L_{1}}(ancd)\,Z_{L_{2}}(cdmv)\,
      Z_{L_{3}}(mwab)\,\langle b||z||n\rangle}{({{\varepsilon }_{mv}}-
       {{\varepsilon }_{cd}})\,
      ({{\varepsilon }_{mw}}-{{\varepsilon }_{ab}})\,
      ({{\varepsilon }_{mnw}}-{{\varepsilon }_{bcd}})}\,+
\]
\[
\frac{1}{[w]}\sum_{abcdmn}\sum_{L_{1}L_{2}L_{3}}
    \frac{\delta _{\kappa }(b,w)\,
       \left\{ \begin{array}[c]{ccc}L_{2} & n &w  \\  d & L_{1} & L_{3}
        \end{array} \right\}\,\left\{ \begin{array}[c]{ccc}L_{3} & L_{2} &
        L_{1}  \\  a & c & m\end{array} \right\}\,X_{L_{1}}(awcd)\,
       X_{L_{2}}(mnab)\,Z_{L_{3}}(cdmn)\,\langle b||z||v\rangle}{
       ({{\varepsilon }_{mn}}-{{\varepsilon }_{ab}})\,
       ({{\varepsilon }_{mn}}-{{\varepsilon }_{cd}})\,
       ({{\varepsilon }_{mnw}}-{{\varepsilon }_{bcd}})}\,+
\]
\[
\,\frac{1}{{\sqrt{[w]}}} \sum_{abcdmn}\sum_{L_{1}}
    \frac{\delta _{\kappa }(a,m)\,\delta _{\kappa }(b,w)\,
       {\left( -1 \right) }^{-a - c + d - n}\,X_{L_{1}}(ancd)\,Z_{0}(mwab)\,
       Z_{L_{1}}(cdmn)\,\langle b||z||v\rangle}{({{\varepsilon }_{mn}}-
        {{\varepsilon }_{cd}})\,
       ({{\varepsilon }_{mw}}-{{\varepsilon }_{ab}})\,
       ({{\varepsilon }_{mnw}}-{{\varepsilon }_{bcd}})\,{\sqrt{[a]}}\,[L_{1}]}
\,+
\]
\[
\sum_{abcdmn}\sum_{L_{1}L_{2}} \frac{{\left( -1 \right) }^
       {b + L_{1} - n - v - w}\,
      \left\{ \begin{array}[c]{ccc}K & L_{2} &L_{1}  \\  a & c & m
       \end{array} \right\}\,\left\{ \begin{array}[c]{ccc}K & v &w  \\  d
        & L_{1} & L_{2}\end{array} \right\}\,Z_{L_{1}}(awcd)\,
      Z_{L_{2}}(bdnv)\,Z_{L_{2}}(mnab)\,\langle c||z||m\rangle}{
      ({{\varepsilon }_{mn}}-{{\varepsilon }_{ab}})\,
      ({{\varepsilon }_{nv}}-{{\varepsilon }_{bd}})\,
      ({{\varepsilon }_{mnw}}-{{\varepsilon }_{bcd}})\,[L_{2}]}\,+
\]
\[
\sum_{abcdmn}\sum_{L_{1}L_{3}} \frac{{\left( -1 \right) }^
       {d + K + L_{1} - n - v - w}\,
      \left\{ \begin{array}[c]{ccc}K & L_{3} &L_{1}  \\  a & c & m
       \end{array} \right\}\,\left\{ \begin{array}[c]{ccc}K & w &v  \\  b
        & L_{1} & L_{3}\end{array} \right\}\,Z_{L_{1}}(ancd)\,
      Z_{L_{1}}(dbnv)\,Z_{L_{3}}(mwab)\,\langle c||z||m\rangle}{
      ({{\varepsilon }_{mw}}-{{\varepsilon }_{ab}})\,
      ({{\varepsilon }_{nv}}-{{\varepsilon }_{bd}})\,
      ({{\varepsilon }_{mnw}}-{{\varepsilon }_{bcd}})\,[L_{1}]}\,+
\]
\[
-\left( \frac{1}{[K]} \right) \sum_{abcdmn}\sum_{L_{2}}
    \frac{{\left( -1 \right) }^{-a + b + c - d + L_{2} - m - n}\,
       \left\{ \begin{array}[c]{ccc}v & w &K  \\  a & d & L_{2}
        \end{array} \right\}\,Z_{K}(andc)\,Z_{L_{2}}(bdmv)\,Z_{L_{2}}(mwba)\,
       \langle c||z||n\rangle}{({{\varepsilon }_{mv}}-{{\varepsilon }_{bd}}
        )\,({{\varepsilon }_{mw}}-{{\varepsilon }_{ab}})\,
       ({{\varepsilon }_{mnw}}-{{\varepsilon }_{bcd}})\,[L_{2}]}\,+
\]
\[
-\left( \frac{1}{[K]} \right) \sum_{abcmnr}\sum_{L_{1}L_{2}}
    \frac{{\left( -1 \right) }^{c - n - v - w}\,
       \left\{ \begin{array}[c]{ccc}K & L_{2} &L_{1}  \\  a & m & r
        \end{array} \right\}\,\left\{ \begin{array}[c]{ccc}K & v &w  \\  b
         & L_{1} & L_{2}\end{array} \right\}\,X_{L_{1}}(mwab)\,
       Z_{K}(nrcm)\,Z_{L_{2}}(abrv)\,\langle c||z||n\rangle}{(
        {{\varepsilon }_{mw}}-{{\varepsilon }_{ab}})\,
       ({{\varepsilon }_{rv}}-{{\varepsilon }_{ab}})\,
       ({{\varepsilon }_{nrw}}-{{\varepsilon }_{abc}})}\,+
\]
\[
-\sum_{abcmnr}\sum_{L_{1}L_{2}L_{3}}
    \frac{{\left( -1 \right) }^{-c + K + L_{3} - v}\,
       \left\{ \begin{array}[c]{ccc}L_{3} & L_{2} &L_{1}  \\  a & m & r
        \end{array} \right\}\,\left\{ \begin{array}[c]{ccc}L_{3} & v &n
          \\  b & L_{1} & L_{2}\end{array} \right\}\,
       \left\{ \begin{array}[c]{ccc}v & L_{3} &n  \\  c & K & w
        \end{array} \right\}\,X_{L_{1}}(mnab)\,Z_{L_{2}}(bavr)\,
       Z_{L_{3}}(rwmc)\,\langle c||z||n\rangle}{({{\varepsilon }_{mn}}-
        {{\varepsilon }_{ab}})\,
       ({{\varepsilon }_{rv}}-{{\varepsilon }_{ab}})\,
       ({{\varepsilon }_{nrw}}-{{\varepsilon }_{abc}})}\,+
\]
\[
-\left( \frac{1}{[K]\,[v]} \right)
   \sum_{abcmnr}\sum_{L_{1}} \frac{\delta _{\kappa }(m,v)\,
       {\left( -1 \right) }^{-a + b + c + K - n - r - v}\,X_{L_{1}}(mnab)\,
       Z_{K}(rwcm)\,Z_{L_{1}}(banv)\,\langle c||z||r\rangle}{(
        {{\varepsilon }_{mn}}-{{\varepsilon }_{ab}})\,
       ({{\varepsilon }_{nv}}-{{\varepsilon }_{ab}})\,
       ({{\varepsilon }_{nrw}}-{{\varepsilon }_{abc}})\,[L_{1}]}\,+
\]
\[
-\left( \frac{1}{[w]} \right) \sum_{abcmnr}\sum_{L_{1}L_{2}L_{3}}
    \frac{\delta _{\kappa }(c,w)\,
       \left\{ \begin{array}[c]{ccc}L_{2} & L_{3} &L_{1}  \\  a & m & r
        \end{array} \right\}\,\left\{ \begin{array}[c]{ccc}L_{2} & n &w
          \\  b & L_{1} & L_{3}\end{array} \right\}\,X_{L_{1}}(mwab)\,
       X_{L_{2}}(nrcm)\,Z_{L_{3}}(banr)\,\langle c||z||v\rangle}{
       ({{\varepsilon }_{mw}}-{{\varepsilon }_{ab}})\,
       ({{\varepsilon }_{nr}}-{{\varepsilon }_{ab}})\,
       ({{\varepsilon }_{nrw}}-{{\varepsilon }_{abc}})}\,+
\]
\[
\frac{1}{[w]}\sum_{abcdmn}\sum_{L_{1}L_{2}}
    \frac{\delta _{\kappa }(a,d)\,\delta _{\kappa }(c,w)\,
       {\left( -1 \right) }^{b + L_{2} + m - n - w}\,X_{L_{1}}(mnab)\,
       Z_{L_{1}}(bdnm)\,Z_{L_{2}}(awcd)\,\langle c||z||v\rangle}{
       ({{\varepsilon }_{mn}}-{{\varepsilon }_{ab}})\,
       ({{\varepsilon }_{mn}}-{{\varepsilon }_{bd}})\,
       ({{\varepsilon }_{mnw}}-{{\varepsilon }_{bcd}})\,[a]\,[L_{1}]}\,+
\]
\[
\frac{1}{[w]}\sum_{abcdmn}\sum_{L_{1}L_{2}}
    \frac{\delta _{\kappa }(c,w)\,
       {\left( -1 \right) }^{-a + b + d + L_{1} - m - n - w}\,
       Z_{L_{1}}(ancd)\,Z_{L_{1}}(mwba)\,Z_{L_{2}}(bdmn)\,
       \langle c||z||v\rangle}{({{\varepsilon }_{mn}}-{{\varepsilon }_{bd}}
        )\,({{\varepsilon }_{mw}}-{{\varepsilon }_{ab}})\,
       ({{\varepsilon }_{mnw}}-{{\varepsilon }_{bcd}})\,{[L_{1}]}^2}\,+
\]
\[
-\left( \frac{1}{[w]} \right) \sum_{abcmnr}\sum_{L_{1}L_{3}}
    \frac{\delta _{\kappa }(c,w)\,\delta _{\kappa }(m,r)\,
       {\left( -1 \right) }^{-a + b + L_{3} - n + w}\,X_{L_{1}}(mnab)\,
       Z_{L_{1}}(banr)\,Z_{L_{3}}(rwcm)\,\langle c||z||v\rangle}{
       ({{\varepsilon }_{mn}}-{{\varepsilon }_{ab}})\,
       ({{\varepsilon }_{nr}}-{{\varepsilon }_{ab}})\,
       ({{\varepsilon }_{nrw}}-{{\varepsilon }_{abc}})\,[L_{1}]\,[m]}\, +
\]
\[
\overline{h.c.s.}
\]

\newpage

\subsection{$Z_{1\times 2}\left( T_v^h \right)$}
\[
Z_{1\times 2}\left( T_v^h \right) =
\]

\[
\sum_{abcmnr}\sum_{L_{1}L_{2}L_{3}}
   \frac{{\left( -1 \right) }^{-a + K + L_{3} + v}\,
      \left\{ \begin{array}[c]{ccc}L_{1} & L_{2} &L_{3}  \\  b & m & n
       \end{array} \right\}\,\left\{ \begin{array}[c]{ccc}L_{1} & v &r
         \\  c & L_{2} & L_{3}\end{array} \right\}\,
      \left\{ \begin{array}[c]{ccc}v & L_{3} &c  \\  a & K & w
       \end{array} \right\}\,X_{L_{1}}(nrmv)\,Z_{L_{2}}(cbrn)\,
      Z_{L_{3}}(wmab)\,\langle a||z||c\rangle}{({{\varepsilon }_{mw}}-
       {{\varepsilon }_{ab}})\,
      ({{\varepsilon }_{nr}}-{{\varepsilon }_{bc}})\,
      ({{\varepsilon }_{nrw}}-{{\varepsilon }_{abv}})}\,+
\]
\[
-\left( \frac{1}{[K]} \right) \sum_{abcmnr}\sum_{L_{1}}
    \frac{\delta _{\kappa }(c,m)\,{\left( -1 \right) }^{a + b + K - n + r}\,
       X_{L_{1}}(nrbm)\,Z_{K}(mwav)\,Z_{L_{1}}(bcnr)\,\langle a||z||c\rangle}
{({{\varepsilon }_{mw}}-{{\varepsilon }_{av}})\,
       ({{\varepsilon }_{nr}}-{{\varepsilon }_{bc}})\,
       ({{\varepsilon }_{nrw}}-{{\varepsilon }_{abv}})\,[c]\,[L_{1}]}\,+
\]
\[
-\sum_{abcmnr}\sum_{L_{1}L_{2}}
    \frac{{\left( -1 \right) }^{b + K + L_{1} - r + v + w}\,
       \left\{ \begin{array}[c]{ccc}L_{1} & L_{2} &K  \\  a & c & n
        \end{array} \right\}\,\left\{ \begin{array}[c]{ccc}m & w &L_{1}
          \\  K & L_{2} & v\end{array} \right\}\,Z_{L_{1}}(bcrn)\,
       Z_{L_{1}}(rwbm)\,Z_{L_{2}}(nmav)\,\langle a||z||c\rangle}{
       ({{\varepsilon }_{mn}}-{{\varepsilon }_{av}})\,
       ({{\varepsilon }_{nr}}-{{\varepsilon }_{bc}})\,
       ({{\varepsilon }_{nrw}}-{{\varepsilon }_{abv}})\,[L_{1}]}\,+
\]
\[
\frac{1}{[K]}\sum_{abcmnr}\sum_{L_{1}}
    \frac{{\left( -1 \right) }^{-a + b - c + L_{1} + m - n - r}\,
       \left\{ \begin{array}[c]{ccc}r & m &K  \\  a & c & L_{1}
        \end{array} \right\}\,Z_{K}(rwmv)\,Z_{L_{1}}(bcnr)\,Z_{L_{1}}(mnab)\,
       \langle a||z||c\rangle}{({{\varepsilon }_{mn}}-{{\varepsilon }_{ab}}
        )\,({{\varepsilon }_{nr}}-{{\varepsilon }_{bc}})\,
       ({{\varepsilon }_{nrw}}-{{\varepsilon }_{abv}})\,[L_{1}]}\,+
\]
\[
\frac{1}{[K]}\sum_{abmnrs}\sum_{L_{1}L_{2}}
    \frac{{\left( -1 \right) }^{a - n - v - w}\,
       \left\{ \begin{array}[c]{ccc}L_{1} & L_{2} &K  \\  b & m & r
        \end{array} \right\}\,\left\{ \begin{array}[c]{ccc}w & v &K  \\
        L_{1} & L_{2} & s\end{array} \right\}\,X_{L_{1}}(rsmv)\,
       Z_{K}(mnba)\,Z_{L_{2}}(bwrs)\,\langle a||z||n\rangle}{(
        {{\varepsilon }_{mn}}-{{\varepsilon }_{ab}})\,
       ({{\varepsilon }_{rs}}-{{\varepsilon }_{bw}})\,
       ({{\varepsilon }_{nrs}}-{{\varepsilon }_{abv}})}\,+
\]
\[
\frac{1}{[K]\,[w]}\sum_{abmnrs}\sum_{L_{1}}
    \frac{\delta _{\kappa }(m,w)\,
       {\left( -1 \right) }^{a + b + K - n - r + s + w}\,X_{L_{1}}(rsbm)\,
       Z_{K}(mnva)\,Z_{L_{1}}(bwrs)\,\langle a||z||n\rangle}{(
        {{\varepsilon }_{mn}}-{{\varepsilon }_{av}})\,
       ({{\varepsilon }_{rs}}-{{\varepsilon }_{bw}})\,
       ({{\varepsilon }_{nrs}}-{{\varepsilon }_{abv}})\,[L_{1}]}\,+
\]
\[
\sum_{abmnrs}\sum_{L_{1}L_{2}} \frac{{\left( -1 \right) }^
       {b + K + L_{1} - s + v + w}\,
      \left\{ \begin{array}[c]{ccc}L_{1} & K &L_{2}  \\  a & m & r
       \end{array} \right\}\,\left\{ \begin{array}[c]{ccc}n & w &L_{1}
         \\  K & L_{2} & v\end{array} \right\}\,Z_{L_{1}}(bwsn)\,
      Z_{L_{1}}(srbm)\,Z_{L_{2}}(mnav)\,\langle a||z||r\rangle}{
      ({{\varepsilon }_{mn}}-{{\varepsilon }_{av}})\,
      ({{\varepsilon }_{ns}}-{{\varepsilon }_{bw}})\,
      ({{\varepsilon }_{nrs}}-{{\varepsilon }_{abv}})\,[L_{1}]}\,+
\]
\[
\sum_{abmnrs}\sum_{L_{1}L_{3}} \frac{{\left( -1 \right) }^
       {b + L_{3} - n + v + w}\,
      \left\{ \begin{array}[c]{ccc}L_{3} & K &L_{1}  \\  a & m & r
       \end{array} \right\}\,\left\{ \begin{array}[c]{ccc}s & v &L_{3}
         \\  K & L_{1} & w\end{array} \right\}\,Z_{L_{1}}(bwns)\,
      Z_{L_{1}}(mnab)\,Z_{L_{3}}(rsmv)\,\langle a||z||r\rangle}{
      ({{\varepsilon }_{mn}}-{{\varepsilon }_{ab}})\,
      ({{\varepsilon }_{ns}}-{{\varepsilon }_{bw}})\,
      ({{\varepsilon }_{nrs}}-{{\varepsilon }_{abv}})\,[L_{1}]}\,+
\]
\[
\frac{{\left( -1 \right) }^{4\,K}}{[K]}
   \sum_{abcmnr}\sum_{L_{1}L_{2}}
    \frac{\left\{ \begin{array}[c]{ccc}L_{1} & L_{2} &K  \\  a & m & r
        \end{array} \right\}\,\left\{ \begin{array}[c]{ccc}L_{2} & L_{1} &
        K  \\  b & c & n\end{array} \right\}\,X_{L_{1}}(nrbm)\,
       Z_{K}(mwav)\,Z_{L_{2}}(acrn)\,\langle b||z||c\rangle}{(
        {{\varepsilon }_{mw}}-{{\varepsilon }_{av}})\,
       ({{\varepsilon }_{nr}}-{{\varepsilon }_{ac}})\,
       ({{\varepsilon }_{nrw}}-{{\varepsilon }_{abv}})}\,+
\]
\[
-\sum_{abcmnr}\sum_{L_{1}L_{3}}
    \frac{{\left( -1 \right) }^{a + L_{3} - n + v + w}\,
       \left\{ \begin{array}[c]{ccc}L_{1} & L_{3} &K  \\  b & c & r
        \end{array} \right\}\,\left\{ \begin{array}[c]{ccc}m & w &L_{3}
          \\  K & L_{1} & v\end{array} \right\}\,Z_{L_{1}}(acnr)\,
       Z_{L_{1}}(nmav)\,Z_{L_{3}}(rwbm)\,\langle b||z||c\rangle}{
       ({{\varepsilon }_{mn}}-{{\varepsilon }_{av}})\,
       ({{\varepsilon }_{nr}}-{{\varepsilon }_{ac}})\,
       ({{\varepsilon }_{nrw}}-{{\varepsilon }_{abv}})\,[L_{1}]}\,+
\]
\[
-\sum_{abmnrs}\sum_{L_{1}L_{2}L_{3}}
    \frac{{\left( -1 \right) }^{K + L_{3} - n + w}\,
       \left\{ \begin{array}[c]{ccc}L_{1} & L_{2} &L_{3}  \\  a & m & s
        \end{array} \right\}\,\left\{ \begin{array}[c]{ccc}L_{2} & w &r
          \\  b & L_{1} & L_{3}\end{array} \right\}\,
       \left\{ \begin{array}[c]{ccc}L_{3} & w &b  \\  K & n & v
        \end{array} \right\}\,X_{L_{1}}(rsbm)\,Z_{L_{2}}(awsr)\,
       Z_{L_{3}}(mnav)\,\langle b||z||n\rangle}{({{\varepsilon }_{mn}}-
        {{\varepsilon }_{av}})\,
       ({{\varepsilon }_{rs}}-{{\varepsilon }_{aw}})\,
       ({{\varepsilon }_{nrs}}-{{\varepsilon }_{abv}})}\,+
\]
\[
-\left( \frac{1}{[K]} \right) \sum_{abmnrs}\sum_{L_{1}}
    \frac{{\left( -1 \right) }^{a + b + L_{1} + m - n - r + s}\,
       \left\{ \begin{array}[c]{ccc}K & w &v  \\  L_{1} & m & s
        \end{array} \right\}\,Z_{K}(rsbm)\,Z_{L_{1}}(awns)\,Z_{L_{1}}(nmav)\,
       \langle b||z||r\rangle}{({{\varepsilon }_{mn}}-{{\varepsilon }_{av}}
        )\,({{\varepsilon }_{ns}}-{{\varepsilon }_{aw}})\,
       ({{\varepsilon }_{nrs}}-{{\varepsilon }_{abv}})\,[L_{1}]}\,+
\]
\[
-\left( \frac{1}{[K]} \right) \sum_{abmnrs}\sum_{L_{1}}
    \frac{{\left( -1 \right) }^{a + b + L_{1} - m + n - r + s}\,
       \left\{ \begin{array}[c]{ccc}n & L_{1} &m  \\  a & K & s
        \end{array} \right\}\,Z_{K}(mwav)\,Z_{L_{1}}(bars)\,Z_{L_{1}}(rnbm)\,
       \langle s||z||n\rangle}{({{\varepsilon }_{mw}}-{{\varepsilon }_{av}}
        )\,({{\varepsilon }_{rs}}-{{\varepsilon }_{ab}})\,
       ({{\varepsilon }_{nrw}}-{{\varepsilon }_{abv}})\,[L_{1}]}\,+
\]
\[
-\sum_{abmnrs}\sum_{L_{1}L_{2}L_{3}}
    \frac{{\left( -1 \right) }^{K + L_{3} - n + w}\,
       \left\{ \begin{array}[c]{ccc}L_{3} & L_{2} &L_{1}  \\  a & m & r
        \end{array} \right\}\,\left\{ \begin{array}[c]{ccc}L_{3} & s &w
          \\  b & L_{1} & L_{2}\end{array} \right\}\,
       \left\{ \begin{array}[c]{ccc}L_{3} & w &s  \\  K & n & v
        \end{array} \right\}\,X_{L_{1}}(mwab)\,Z_{L_{2}}(basr)\,
       Z_{L_{3}}(rnmv)\,\langle s||z||n\rangle}{({{\varepsilon }_{mw}}-
        {{\varepsilon }_{ab}})\,
       ({{\varepsilon }_{rs}}-{{\varepsilon }_{ab}})\,
       ({{\varepsilon }_{nrw}}-{{\varepsilon }_{abv}})}\,+
\]
\[
-\sum_{abmnrs}\sum_{L_{1}L_{2}}
    \frac{{\left( -1 \right) }^{-a + b + K + L_{1} + L_{2} - m - n - r + s}\,
       \left\{ \begin{array}[c]{ccc}L_{2} & w &s  \\  K & n & v
        \end{array} \right\}\,\left\{ \begin{array}[c]{ccc}w & L_{2} &s
          \\  a & L_{1} & m\end{array} \right\}\,Z_{L_{1}}(bars)\,
       Z_{L_{1}}(rwbm)\,Z_{L_{2}}(mnav)\,\langle s||z||n\rangle}{
       ({{\varepsilon }_{mn}}-{{\varepsilon }_{av}})\,
       ({{\varepsilon }_{rs}}-{{\varepsilon }_{ab}})\,
       ({{\varepsilon }_{nrw}}-{{\varepsilon }_{abv}})\,[L_{1}]}\,+
\]
\[
-\left( \frac{1}{[K]} \right) \sum_{abmnrs}\sum_{L_{1}L_{2}}
    \frac{\left\{ \begin{array}[c]{ccc}K & L_{2} &L_{1}  \\  a & m & r
        \end{array} \right\}\,\left\{ \begin{array}[c]{ccc}K & s &n  \\  b
         & L_{1} & L_{2}\end{array} \right\}\,X_{L_{1}}(mnab)\,
       Z_{K}(rwmv)\,Z_{L_{2}}(abrs)\,\langle s||z||n\rangle}{(
        {{\varepsilon }_{mn}}-{{\varepsilon }_{ab}})\,
       ({{\varepsilon }_{rs}}-{{\varepsilon }_{ab}})\,
       ({{\varepsilon }_{nrw}}-{{\varepsilon }_{abv}})}\,+
\]
\[
\sum_{abmnrs}\sum_{L_{1}L_{3}} \frac{{\left( -1 \right) }^
       {a + K + L_{1} - n + v + w}\,
      \left\{ \begin{array}[c]{ccc}K & L_{1} &L_{3}  \\  b & r & s
       \end{array} \right\}\,\left\{ \begin{array}[c]{ccc}m & w &L_{3}
         \\  K & L_{1} & v\end{array} \right\}\,Z_{L_{1}}(abns)\,
      Z_{L_{1}}(nmav)\,Z_{L_{3}}(rwbm)\,\langle s||z||r\rangle}{
      ({{\varepsilon }_{mn}}-{{\varepsilon }_{av}})\,
      ({{\varepsilon }_{ns}}-{{\varepsilon }_{ab}})\,
      ({{\varepsilon }_{nrw}}-{{\varepsilon }_{abv}})\,[L_{1}]}\,+
\]
\[
-\left( \frac{1}{[K]} \right) \sum_{abmnrs}\sum_{L_{1}}
    \frac{\delta _{\kappa }(m,s)\,{\left( -1 \right) }^{-a + b + K - n + r}\,
       X_{L_{1}}(mnab)\,Z_{K}(rwmv)\,Z_{L_{1}}(bans)\,\langle s||z||r\rangle}
{({{\varepsilon }_{mn}}-{{\varepsilon }_{ab}})\,
       ({{\varepsilon }_{ns}}-{{\varepsilon }_{ab}})\,
       ({{\varepsilon }_{nrw}}-{{\varepsilon }_{abv}})\,[L_{1}]\,[m]}\,+
\]
\[
\frac{1}{[v]}\sum_{abmnrs}\sum_{L_{1}L_{2}L_{3}}
    \frac{\delta _{\kappa }(n,v)\,
       \left\{ \begin{array}[c]{ccc}L_{2} & L_{3} &L_{1}  \\  a & m & r
        \end{array} \right\}\,\left\{ \begin{array}[c]{ccc}L_{2} & s &v
          \\  b & L_{1} & L_{3}\end{array} \right\}\,X_{L_{1}}(mnab)\,
       X_{L_{2}}(rsmv)\,Z_{L_{3}}(abrs)\,\langle w||z||n\rangle}{
       ({{\varepsilon }_{mn}}-{{\varepsilon }_{ab}})\,
       ({{\varepsilon }_{rs}}-{{\varepsilon }_{ab}})\,
       ({{\varepsilon }_{nrs}}-{{\varepsilon }_{abv}})}\,+
\]
\[
\,\frac{1}{{\sqrt{[v]}}} \sum_{abmnrs}\sum_{L_{1}}
    \frac{\delta _{\kappa }(a,m)\,\delta _{\kappa }(n,v)\,
       {\left( -1 \right) }^{a + b - r + s}\,X_{L_{1}}(rsbm)\,Z_{0}(mnav)\,
       Z_{L_{1}}(bars)\,\langle w||z||n\rangle}{({{\varepsilon }_{mn}}-
        {{\varepsilon }_{av}})\,
       ({{\varepsilon }_{rs}}-{{\varepsilon }_{ab}})\,
       ({{\varepsilon }_{nrs}}-{{\varepsilon }_{abv}})\,{\sqrt{[a]}}\,[L_{1}]}
\,+
\]
\[
\frac{1}{[v]}\sum_{abmnrs}\sum_{L_{1}L_{2}}
    \frac{\delta _{\kappa }(r,v)\,
       {\left( -1 \right) }^{a + b + L_{2} + m - n - s + v}\,
       Z_{L_{1}}(abns)\,Z_{L_{1}}(nmav)\,Z_{L_{2}}(srbm)\,
       \langle w||z||r\rangle}{({{\varepsilon }_{mn}}-{{\varepsilon }_{av}}
        )\,({{\varepsilon }_{ns}}-{{\varepsilon }_{ab}})\,
       ({{\varepsilon }_{nrs}}-{{\varepsilon }_{abv}})\,{[L_{2}]}^2}\,+
\]
\[
\frac{1}{[v]}\sum_{abmnrs}\sum_{L_{1}L_{3}}
    \frac{\delta _{\kappa }(m,s)\,\delta _{\kappa }(r,v)\,
       {\left( -1 \right) }^{-a + b + L_{3} - n + v}\,X_{L_{1}}(mnab)\,
       Z_{L_{1}}(bans)\,Z_{L_{3}}(rsmv)\,\langle w||z||r\rangle}{
       ({{\varepsilon }_{mn}}-{{\varepsilon }_{ab}})\,
       ({{\varepsilon }_{ns}}-{{\varepsilon }_{ab}})\,
       ({{\varepsilon }_{nrs}}-{{\varepsilon }_{abv}})\,[L_{1}]\,[m]}\, +
\]
\[
\overline{h.c.s.}
\]

\newpage

\subsection{$Z_{1\times 2}\left( T_v^p \right)$}
\[
Z_{1\times 2}\left( T_v^p \right) =
\]

\[
\sum_{abcdmn}\sum_{L_{1}L_{2}L_{3}}
   \frac{{\left( -1 \right) }^{-b + K + L_{2} - v}\,
      \left\{ \begin{array}[c]{ccc}L_{1} & n &v  \\  d & L_{2} & L_{3}
       \end{array} \right\}\,\left\{ \begin{array}[c]{ccc}L_{3} & L_{1} &
       L_{2}  \\  a & c & m\end{array} \right\}\,
      \left\{ \begin{array}[c]{ccc}v & L_{2} &d  \\  b & K & w
       \end{array} \right\}\,X_{L_{1}}(mnav)\,Z_{L_{2}}(awcb)\,
      Z_{L_{3}}(cdmn)\,\langle b||z||d\rangle}{({{\varepsilon }_{mn}}-
       {{\varepsilon }_{av}})\,
      ({{\varepsilon }_{mn}}-{{\varepsilon }_{cd}})\,
      ({{\varepsilon }_{mnw}}-{{\varepsilon }_{bcv}})}\,+
\]
\[
-\left( \frac{1}{[K]} \right) \sum_{abcdmn}\sum_{L_{1}L_{3}}
    \frac{\left\{ \begin{array}[c]{ccc}L_{3} & L_{1} &K  \\  a & c & m
        \end{array} \right\}\,\left\{ \begin{array}[c]{ccc}L_{3} & L_{1} &
        K  \\  b & d & n\end{array} \right\}\,X_{L_{1}}(mnab)\,
       Z_{K}(awcv)\,Z_{L_{3}}(cdmn)\,\langle b||z||d\rangle}{(
        {{\varepsilon }_{mn}}-{{\varepsilon }_{ab}})\,
       ({{\varepsilon }_{mn}}-{{\varepsilon }_{cd}})\,
       ({{\varepsilon }_{mnw}}-{{\varepsilon }_{bcv}})}\,+
\]
\[
\sum_{abcdmn}\sum_{L_{1}L_{2}} \frac{{\left( -1 \right) }^
       {-a - b + c - d + K + L_{1} + L_{2} - m - n}\,
      \left\{ \begin{array}[c]{ccc}d & L_{2} &v  \\  a & L_{1} & m
       \end{array} \right\}\,\left\{ \begin{array}[c]{ccc}v & w &K  \\  b
        & d & L_{2}\end{array} \right\}\,Z_{L_{1}}(cdnm)\,Z_{L_{1}}(nacv)\,
      Z_{L_{2}}(mwab)\,\langle b||z||d\rangle}{({{\varepsilon }_{mn}}-
       {{\varepsilon }_{cd}})\,
      ({{\varepsilon }_{mw}}-{{\varepsilon }_{ab}})\,
      ({{\varepsilon }_{mnw}}-{{\varepsilon }_{bcv}})\,[L_{1}]}\,+
\]
\[
-\left( \frac{1}{[K]} \right) \sum_{abcdmn}\sum_{L_{1}}
    \frac{{\left( -1 \right) }^{-a - b + c - d + L_{1} + m - n}\,
       \left\{ \begin{array}[c]{ccc}d & m &L_{1}  \\  a & b & K
        \end{array} \right\}\,Z_{K}(mwav)\,Z_{L_{1}}(anbc)\,Z_{L_{1}}(cdnm)\,
       \langle b||z||d\rangle}{({{\varepsilon }_{mn}}-{{\varepsilon }_{cd}}
        )\,({{\varepsilon }_{mw}}-{{\varepsilon }_{av}})\,
       ({{\varepsilon }_{mnw}}-{{\varepsilon }_{bcv}})\,[L_{1}]}\,+
\]
\[
-\left( \frac{1}{[K]} \right) \sum_{abcmnr}\sum_{L_{1}}
    \frac{{\left( -1 \right) }^{a + b + c + L_{1} - m - n - r}\,
       \left\{ \begin{array}[c]{ccc}w & L_{1} &n  \\  a & K & v
        \end{array} \right\}\,Z_{K}(mnba)\,Z_{L_{1}}(cwrn)\,Z_{L_{1}}(racv)\,
       \langle b||z||m\rangle}{({{\varepsilon }_{mn}}-{{\varepsilon }_{ab}}
        )\,({{\varepsilon }_{nr}}-{{\varepsilon }_{cw}})\,
       ({{\varepsilon }_{mnr}}-{{\varepsilon }_{bcv}})\,[L_{1}]}\,+
\]
\[
-\sum_{abcmnr}\sum_{L_{1}L_{3}}
    \frac{{\left( -1 \right) }^{c + L_{3} - r + v + w}\,
       \left\{ \begin{array}[c]{ccc}K & L_{3} &L_{1}  \\  a & b & m
        \end{array} \right\}\,\left\{ \begin{array}[c]{ccc}n & v &L_{3}
          \\  K & L_{1} & w\end{array} \right\}\,Z_{L_{1}}(arbc)\,
       Z_{L_{1}}(cwrn)\,Z_{L_{3}}(mnav)\,\langle b||z||m\rangle}{
       ({{\varepsilon }_{mn}}-{{\varepsilon }_{av}})\,
       ({{\varepsilon }_{nr}}-{{\varepsilon }_{cw}})\,
       ({{\varepsilon }_{mnr}}-{{\varepsilon }_{bcv}})\,[L_{1}]}\,+
\]
\[
-\left( \frac{1}{[K]} \right) \sum_{abcmnr}\sum_{L_{1}L_{3}}
    \frac{{\left( -1 \right) }^{b - r - v - w}\,
       \left\{ \begin{array}[c]{ccc}L_{3} & L_{1} &K  \\  a & c & m
        \end{array} \right\}\,\left\{ \begin{array}[c]{ccc}w & n &L_{3}
          \\  L_{1} & K & v\end{array} \right\}\,X_{L_{1}}(mnav)\,
       Z_{K}(arcb)\,Z_{L_{3}}(cwmn)\,\langle b||z||r\rangle}{(
        {{\varepsilon }_{mn}}-{{\varepsilon }_{av}})\,
       ({{\varepsilon }_{mn}}-{{\varepsilon }_{cw}})\,
       ({{\varepsilon }_{mnr}}-{{\varepsilon }_{bcv}})}\,+
\]
\[
\sum_{abcmnr}\sum_{L_{1}L_{2}L_{3}}
   \frac{{\left( -1 \right) }^{K + L_{2} - r + w}\,
      \left\{ \begin{array}[c]{ccc}L_{2} & w &b  \\  K & r & v
       \end{array} \right\}\,\left\{ \begin{array}[c]{ccc}L_{3} & L_{1} &
       L_{2}  \\  a & c & m\end{array} \right\}\,
      \left\{ \begin{array}[c]{ccc}L_{3} & w &n  \\  b & L_{1} & L_{2}
       \end{array} \right\}\,X_{L_{1}}(mnab)\,Z_{L_{2}}(arcv)\,
      Z_{L_{3}}(cwmn)\,\langle b||z||r\rangle}{({{\varepsilon }_{mn}}-
       {{\varepsilon }_{ab}})\,
      ({{\varepsilon }_{mn}}-{{\varepsilon }_{cw}})\,
      ({{\varepsilon }_{mnr}}-{{\varepsilon }_{bcv}})}\,+
\]
\[
\frac{1}{[K]}\sum_{abcdmn}\sum_{L_{1}}
    \frac{\delta _{\kappa }(a,d)\,{\left( -1 \right) }^{b + c + K + m - n}\,
       X_{L_{1}}(mnab)\,Z_{K}(awcv)\,Z_{L_{1}}(bdnm)\,\langle c||z||d\rangle}
{({{\varepsilon }_{mn}}-{{\varepsilon }_{ab}})\,
       ({{\varepsilon }_{mn}}-{{\varepsilon }_{bd}})\,
       ({{\varepsilon }_{mnw}}-{{\varepsilon }_{bcv}})\,[a]\,[L_{1}]}\,+
\]
\[
-\sum_{abcdmn}\sum_{L_{1}L_{3}}
    \frac{{\left( -1 \right) }^{b + K + L_{1} - m + v - w}\,
       \left\{ \begin{array}[c]{ccc}K & L_{1} &L_{3}  \\  a & v & w
        \end{array} \right\}\,\left\{ \begin{array}[c]{ccc}L_{1} & L_{3} &
        K  \\  c & d & n\end{array} \right\}\,Z_{L_{1}}(bdmn)\,
       Z_{L_{1}}(mwba)\,Z_{L_{3}}(nacv)\,\langle c||z||d\rangle}{
       ({{\varepsilon }_{mn}}-{{\varepsilon }_{bd}})\,
       ({{\varepsilon }_{mw}}-{{\varepsilon }_{ab}})\,
       ({{\varepsilon }_{mnw}}-{{\varepsilon }_{bcv}})\,[L_{1}]}\,+
\]
\[
-\sum_{abcmnr}\sum_{L_{1}L_{2}}
    \frac{{\left( -1 \right) }^{b + K + L_{2} - n + v + w}\,
       \left\{ \begin{array}[c]{ccc}K & L_{2} &L_{1}  \\  a & c & m
        \end{array} \right\}\,\left\{ \begin{array}[c]{ccc}r & w &L_{2}
          \\  K & L_{1} & v\end{array} \right\}\,Z_{L_{1}}(arcv)\,
       Z_{L_{2}}(bwnr)\,Z_{L_{2}}(mnab)\,\langle c||z||m\rangle}{
       ({{\varepsilon }_{mn}}-{{\varepsilon }_{ab}})\,
       ({{\varepsilon }_{nr}}-{{\varepsilon }_{bw}})\,
       ({{\varepsilon }_{mnr}}-{{\varepsilon }_{bcv}})\,[L_{2}]}\,+
\]
\[
-\left( \frac{1}{[K]\,[w]} \right)
   \sum_{abcmnr}\sum_{L_{1}} \frac{\delta _{\kappa }(a,w)\,
       {\left( -1 \right) }^{b + c + K + m - n - r + w}\,X_{L_{1}}(mnab)\,
       Z_{K}(arvc)\,Z_{L_{1}}(bwnm)\,\langle c||z||r\rangle}{(
        {{\varepsilon }_{mn}}-{{\varepsilon }_{ab}})\,
       ({{\varepsilon }_{mn}}-{{\varepsilon }_{bw}})\,
       ({{\varepsilon }_{mnr}}-{{\varepsilon }_{bcv}})\,[L_{1}]}\,+
\]
\[
\frac{1}{[K]}\sum_{abcmnr}\sum_{L_{2}}
    \frac{{\left( -1 \right) }^{-a + b - c + L_{2} + m - n - r}\,
       \left\{ \begin{array}[c]{ccc}r & m &K  \\  a & c & L_{2}
        \end{array} \right\}\,Z_{K}(awcv)\,Z_{L_{2}}(bcnr)\,Z_{L_{2}}(mnab)\,
       \langle r||z||m\rangle}{({{\varepsilon }_{mn}}-{{\varepsilon }_{ab}}
        )\,({{\varepsilon }_{nr}}-{{\varepsilon }_{bc}})\,
       ({{\varepsilon }_{mnw}}-{{\varepsilon }_{bcv}})\,[L_{2}]}\,+
\]
\[
\sum_{abcmnr}\sum_{L_{1}L_{2}L_{3}}
   \frac{{\left( -1 \right) }^{K + L_{3} - m + w}\,
      \left\{ \begin{array}[c]{ccc}L_{2} & L_{3} &L_{1}  \\  a & b & n
       \end{array} \right\}\,\left\{ \begin{array}[c]{ccc}L_{3} & r &w
         \\  c & L_{1} & L_{2}\end{array} \right\}\,
      \left\{ \begin{array}[c]{ccc}L_{3} & w &r  \\  K & m & v
       \end{array} \right\}\,X_{L_{1}}(awbc)\,Z_{L_{2}}(cbrn)\,
      Z_{L_{3}}(nmav)\,\langle r||z||m\rangle}{({{\varepsilon }_{mn}}-
       {{\varepsilon }_{av}})\,
      ({{\varepsilon }_{nr}}-{{\varepsilon }_{bc}})\,
      ({{\varepsilon }_{mnw}}-{{\varepsilon }_{bcv}})}\,+
\]
\[
\sum_{abcmnr}\sum_{L_{1}L_{2}} \frac{{\left( -1 \right) }^
       {c + K + L_{1} - n + v - w}\,
      \left\{ \begin{array}[c]{ccc}K & L_{2} &L_{1}  \\  a & v & w
       \end{array} \right\}\,\left\{ \begin{array}[c]{ccc}K & r &m  \\  b
        & L_{2} & L_{1}\end{array} \right\}\,Z_{L_{1}}(cbnr)\,
      Z_{L_{1}}(nacv)\,Z_{L_{2}}(mwba)\,\langle r||z||m\rangle}{
      ({{\varepsilon }_{mw}}-{{\varepsilon }_{ab}})\,
      ({{\varepsilon }_{nr}}-{{\varepsilon }_{bc}})\,
      ({{\varepsilon }_{mnw}}-{{\varepsilon }_{bcv}})\,[L_{1}]}\,+
\]
\[
\frac{1}{[K]}\sum_{abcmnr}\sum_{L_{1}}
    \frac{\delta _{\kappa }(a,r)\,{\left( -1 \right) }^{-b + c + K + m - n}\,
       X_{L_{1}}(anbc)\,Z_{K}(mwav)\,Z_{L_{1}}(cbnr)\,\langle r||z||m\rangle}
{({{\varepsilon }_{mw}}-{{\varepsilon }_{av}})\,
       ({{\varepsilon }_{nr}}-{{\varepsilon }_{bc}})\,
       ({{\varepsilon }_{mnw}}-{{\varepsilon }_{bcv}})\,[a]\,[L_{1}]}\,+
\]
\[
\sum_{abcmnr}\sum_{L_{1}L_{2}} \frac{{\left( -1 \right) }^
       {-a + b - c + K + L_{1} + L_{2} - m - n + r}\,
      \left\{ \begin{array}[c]{ccc}L_{1} & w &r  \\  K & n & v
       \end{array} \right\}\,\left\{ \begin{array}[c]{ccc}r & w &L_{1}
         \\  a & c & L_{2}\end{array} \right\}\,Z_{L_{1}}(ancv)\,
      Z_{L_{2}}(bcmr)\,Z_{L_{2}}(mwba)\,\langle r||z||n\rangle}{
      ({{\varepsilon }_{mr}}-{{\varepsilon }_{bc}})\,
      ({{\varepsilon }_{mw}}-{{\varepsilon }_{ab}})\,
      ({{\varepsilon }_{mnw}}-{{\varepsilon }_{bcv}})\,[L_{2}]}\,+
\]
\[
\frac{1}{[K]}\sum_{abcmnr}\sum_{L_{1}L_{2}}
    \frac{\left\{ \begin{array}[c]{ccc}K & r &n  \\  c & L_{1} & L_{2}
        \end{array} \right\}\,\left\{ \begin{array}[c]{ccc}L_{2} & K &
        L_{1}  \\  a & b & m\end{array} \right\}\,X_{L_{1}}(anbc)\,
       Z_{K}(mwav)\,Z_{L_{2}}(bcmr)\,\langle r||z||n\rangle}{(
        {{\varepsilon }_{mr}}-{{\varepsilon }_{bc}})\,
       ({{\varepsilon }_{mw}}-{{\varepsilon }_{av}})\,
       ({{\varepsilon }_{mnw}}-{{\varepsilon }_{bcv}})}\,+
\]
\[
-\left( \frac{1}{[v]} \right) \sum_{abcmnr}\sum_{L_{1}L_{2}}
    \frac{\delta _{\kappa }(m,v)\,
       {\left( -1 \right) }^{-a + b + c + L_{2} - n - r - v}\,
       Z_{L_{1}}(bcnr)\,Z_{L_{1}}(mnab)\,Z_{L_{2}}(racv)\,
       \langle w||z||m\rangle}{({{\varepsilon }_{mn}}-{{\varepsilon }_{ab}}
        )\,({{\varepsilon }_{nr}}-{{\varepsilon }_{bc}})\,
       ({{\varepsilon }_{mnr}}-{{\varepsilon }_{bcv}})\,{[L_{2}]}^2}\,+
\]
\[
-\left( \frac{1}{[v]} \right) \sum_{abcmnr}\sum_{L_{1}L_{3}}
    \frac{\delta _{\kappa }(a,n)\,\delta _{\kappa }(m,v)\,
       {\left( -1 \right) }^{-b + c + L_{3} - r + v}\,X_{L_{1}}(arbc)\,
       Z_{L_{1}}(bcnr)\,Z_{L_{3}}(mnav)\,\langle w||z||m\rangle}{
       ({{\varepsilon }_{mn}}-{{\varepsilon }_{av}})\,
       ({{\varepsilon }_{nr}}-{{\varepsilon }_{bc}})\,
       ({{\varepsilon }_{mnr}}-{{\varepsilon }_{bcv}})\,[a]\,[L_{1}]}\,+
\]
\[
-\left( \frac{1}{[v]} \right) \sum_{abcmnr}\sum_{L_{1}L_{2}L_{3}}
    \frac{\delta _{\kappa }(r,v)\,
       \left\{ \begin{array}[c]{ccc}L_{2} & n &v  \\  c & L_{1} & L_{3}
        \end{array} \right\}\,\left\{ \begin{array}[c]{ccc}L_{3} & L_{2} &
        L_{1}  \\  a & b & m\end{array} \right\}\,X_{L_{1}}(arbc)\,
       X_{L_{2}}(mnav)\,Z_{L_{3}}(bcmn)\,\langle w||z||r\rangle}{
       ({{\varepsilon }_{mn}}-{{\varepsilon }_{av}})\,
       ({{\varepsilon }_{mn}}-{{\varepsilon }_{bc}})\,
       ({{\varepsilon }_{mnr}}-{{\varepsilon }_{bcv}})}\,+
\]
\[
-\,\frac{1}{{\sqrt{[v]}}} \sum_{abcmnr}\sum_{L_{1}}
     \frac{\delta _{\kappa }(a,c)\,\delta _{\kappa }(r,v)\,
        {\left( -1 \right) }^{a + b + m - n}\,X_{L_{1}}(mnab)\,Z_{0}(arcv)\,
        Z_{L_{1}}(bcnm)\,\langle w||z||r\rangle}{({{\varepsilon }_{mn}}-
         {{\varepsilon }_{ab}})\,
        ({{\varepsilon }_{mn}}-{{\varepsilon }_{bc}})\,
        ({{\varepsilon }_{mnr}}-{{\varepsilon }_{bcv}})\,{\sqrt{[a]}}\,
        [L_{1}]}\, +
\]
\[
\overline{h.c.s.}
\]

\newpage


\subsection{$Z_{0\times 3}\left( S_v[T_v] \right)$}
\[
Z_{0\times 3}\left( S_v[T_v] \right) =
\]

\[
\frac{1}{[v]}\sum_{abmnrs}\sum_{L_{1}L_{2}L_{3}}
    \frac{\delta _{\kappa }(n,v)\,
       \left\{ \begin{array}[c]{ccc}L_{2} & L_{3} &L_{1}  \\  a & m & r
        \end{array} \right\}\,\left\{ \begin{array}[c]{ccc}L_{2} & s &v
          \\  b & L_{1} & L_{3}\end{array} \right\}\,X_{L_{1}}(mnab)\,
       X_{L_{2}}(rsmv)\,Z_{L_{3}}(abrs)\,\langle w||z||n\rangle}{
       \left( {{\varepsilon }_n} - {{\varepsilon }_{v}} \right) \,
       ({{\varepsilon }_{mn}}-{{\varepsilon }_{ab}})\,
       ({{\varepsilon }_{nrs}}-{{\varepsilon }_{abv}})}\,+
\]
\[
-\left( \frac{1}{[v]} \right) \sum_{abcmnr}\sum_{L_{1}L_{2}}
    \frac{\delta _{\kappa }(n,v)\,
       {\left( -1 \right) }^{-a + b + c + L_{2} - m - r - v}\,
       Z_{L_{1}}(bcmr)\,Z_{L_{1}}(mnba)\,Z_{L_{2}}(racv)\,
       \langle w||z||n\rangle}{\left( {{\varepsilon }_n} -
         {{\varepsilon }_{v}} \right) \,
       ({{\varepsilon }_{mn}}-{{\varepsilon }_{ab}})\,
       ({{\varepsilon }_{mnr}}-{{\varepsilon }_{bcv}})\,{[L_{2}]}^2}\,+
\]
\[
\,\frac{1}{{\sqrt{[v]}}} \sum_{abmnrs}\sum_{L_{1}}
    \frac{\delta _{\kappa }(a,m)\,\delta _{\kappa }(n,v)\,
       {\left( -1 \right) }^{a + b - r + s}\,X_{L_{1}}(rsbm)\,Z_{0}(mnav)\,
       Z_{L_{1}}(bars)\,\langle w||z||n\rangle}{\left( {{\varepsilon }_n} -
         {{\varepsilon }_{v}} \right) \,
       ({{\varepsilon }_{mn}}-{{\varepsilon }_{av}})\,
       ({{\varepsilon }_{nrs}}-{{\varepsilon }_{abv}})\,{\sqrt{[a]}}\,[L_{1}]}
\,+
\]
\[
-\,\frac{1}{{\sqrt{[v]}}} \sum_{abcmnr}\sum_{L_{1}}
     \frac{\delta _{\kappa }(a,m)\,\delta _{\kappa }(n,v)\,
        {\left( -1 \right) }^{-a - b + c - r}\,X_{L_{1}}(arbc)\,Z_{0}(mnav)\,
        Z_{L_{1}}(bcmr)\,\langle w||z||n\rangle}{\left( {{\varepsilon }_
             n} - {{\varepsilon }_{v}} \right) \,
        ({{\varepsilon }_{mn}}-{{\varepsilon }_{av}})\,
        ({{\varepsilon }_{mnr}}-{{\varepsilon }_{bcv}})\,{\sqrt{[a]}}\,
        [L_{1}]}\,+
\]
\[
-\left( \frac{1}{[v]} \right) \sum_{abcmnr}\sum_{L_{1}L_{2}L_{3}}
    \frac{\delta _{\kappa }(r,v)\,
       \left\{ \begin{array}[c]{ccc}L_{2} & n &v  \\  c & L_{1} & L_{3}
        \end{array} \right\}\,\left\{ \begin{array}[c]{ccc}L_{3} & L_{2} &
        L_{1}  \\  a & b & m\end{array} \right\}\,X_{L_{1}}(arbc)\,
       X_{L_{2}}(mnav)\,Z_{L_{3}}(bcmn)\,\langle w||z||r\rangle}{
       \left( {{\varepsilon }_r} - {{\varepsilon }_{v}} \right) \,
       ({{\varepsilon }_{mn}}-{{\varepsilon }_{av}})\,
       ({{\varepsilon }_{mnr}}-{{\varepsilon }_{bcv}})}\,+
\]
\[
-\,\frac{1}{{\sqrt{[v]}}} \sum_{abcmnr}\sum_{L_{1}}
     \frac{\delta _{\kappa }(a,c)\,\delta _{\kappa }(r,v)\,
        {\left( -1 \right) }^{a + b + m - n}\,X_{L_{1}}(mnab)\,Z_{0}(arcv)\,
        Z_{L_{1}}(bcnm)\,\langle w||z||r\rangle}{\left( {{\varepsilon }_
             r} - {{\varepsilon }_{v}} \right) \,
        ({{\varepsilon }_{mn}}-{{\varepsilon }_{ab}})\,
        ({{\varepsilon }_{mnr}}-{{\varepsilon }_{bcv}})\,{\sqrt{[a]}}\,
        [L_{1}]}\,+
\]
\[
\frac{1}{[v]}\sum_{abmnrs}\sum_{L_{1}L_{2}}
    \frac{\delta _{\kappa }(s,v)\,
       {\left( -1 \right) }^{a + b + L_{2} + m - n - r + v}\,
       Z_{L_{1}}(abnr)\,Z_{L_{1}}(nmav)\,Z_{L_{2}}(rsbm)\,
       \langle w||z||s\rangle}{\left( {{\varepsilon }_s} -
         {{\varepsilon }_{v}} \right) \,
       ({{\varepsilon }_{mn}}-{{\varepsilon }_{av}})\,
       ({{\varepsilon }_{nrs}}-{{\varepsilon }_{abv}})\,{[L_{2}]}^2}\,+
\]
\[
\,\frac{1}{{\sqrt{[v]}}} \sum_{abmnrs}\sum_{L_{1}}
    \frac{\delta _{\kappa }(m,r)\,\delta _{\kappa }(s,v)\,
       {\left( -1 \right) }^{-a + b - m - n}\,X_{L_{1}}(mnab)\,Z_{0}(rsmv)\,
       Z_{L_{1}}(banr)\,\langle w||z||s\rangle}{\left( {{\varepsilon }_s} -
         {{\varepsilon }_{v}} \right) \,
       ({{\varepsilon }_{mn}}-{{\varepsilon }_{ab}})\,
       ({{\varepsilon }_{nrs}}-{{\varepsilon }_{abv}})\,[L_{1}]\,{\sqrt{[m]}}}
\, + 
\]
\[
\overline{h.c.s.}
\]

\newpage

\subsection{$Z_{0\times 3}\left( S_c[T_c] \right)$}
\[
Z_{0\times 3}\left( S_c[T_c] \right) =
\]

\[
\frac{1}{[w]}\sum_{abcdmn}\sum_{L_{1}L_{2}L_{3}}
    \frac{\delta _{\kappa }(b,w)\,
       \left\{ \begin{array}[c]{ccc}L_{2} & n &w  \\  d & L_{1} & L_{3}
        \end{array} \right\}\,\left\{ \begin{array}[c]{ccc}L_{3} & L_{2} &
        L_{1}  \\  a & c & m\end{array} \right\}\,X_{L_{1}}(awcd)\,
       X_{L_{2}}(mnab)\,Z_{L_{3}}(cdmn)\,\langle b||z||v\rangle}{
       ({{\varepsilon }_w}-{{\varepsilon }_{b}})\,
       ({{\varepsilon }_{mn}}-{{\varepsilon }_{ab}})\,
       ({{\varepsilon }_{mnw}}-{{\varepsilon }_{bcd}})}\,+
\]
\[
-\,\frac{1}{{\sqrt{[w]}}} \sum_{abcmnr}\sum_{L_{1}}
     \frac{\delta _{\kappa }(a,m)\,\delta _{\kappa }(b,w)\,
        {\left( -1 \right) }^{a + c - n + r}\,X_{L_{1}}(nrcm)\,Z_{0}(mwab)\,
        Z_{L_{1}}(canr)\,\langle b||z||v\rangle}{({{\varepsilon }_w}-
         {{\varepsilon }_{b}})\,
        ({{\varepsilon }_{mw}}-{{\varepsilon }_{ab}})\,
        ({{\varepsilon }_{nrw}}-{{\varepsilon }_{abc}})\,{\sqrt{[a]}}\,
        [L_{1}]}\,+
\]
\[
\,\frac{1}{{\sqrt{[w]}}} \sum_{abcdmn}\sum_{L_{1}}
    \frac{\delta _{\kappa }(a,m)\,\delta _{\kappa }(b,w)\,
       {\left( -1 \right) }^{-a - c + d - n}\,X_{L_{1}}(ancd)\,Z_{0}(mwab)\,
       Z_{L_{1}}(cdmn)\,\langle b||z||v\rangle}{({{\varepsilon }_w}-
        {{\varepsilon }_{b}})\,
       ({{\varepsilon }_{mw}}-{{\varepsilon }_{ab}})\,
       ({{\varepsilon }_{mnw}}-{{\varepsilon }_{bcd}})\,{\sqrt{[a]}}\,[L_{1}]}
\,+
\]
\[
-\left( \frac{1}{[w]} \right) \sum_{abcmnr}\sum_{L_{1}L_{2}}
    \frac{\delta _{\kappa }(b,w)\,
       {\left( -1 \right) }^{a + c + L_{2} + m - n - r + w}\,
       Z_{L_{1}}(acnr)\,Z_{L_{1}}(nmab)\,Z_{L_{2}}(rwcm)\,
       \langle b||z||v\rangle}{({{\varepsilon }_w}-{{\varepsilon }_{b}})\,
       ({{\varepsilon }_{mn}}-{{\varepsilon }_{ab}})\,
       ({{\varepsilon }_{nrw}}-{{\varepsilon }_{abc}})\,{[L_{2}]}^2}\,+
\]
\[
-\left( \frac{1}{[w]} \right) \sum_{abcmnr}\sum_{L_{1}L_{2}L_{3}}
    \frac{\delta _{\kappa }(c,w)\,
       \left\{ \begin{array}[c]{ccc}L_{2} & L_{3} &L_{1}  \\  a & m & r
        \end{array} \right\}\,\left\{ \begin{array}[c]{ccc}L_{2} & n &w
          \\  b & L_{1} & L_{3}\end{array} \right\}\,X_{L_{1}}(mwab)\,
       X_{L_{2}}(nrcm)\,Z_{L_{3}}(banr)\,\langle c||z||v\rangle}{
       ({{\varepsilon }_w}-{{\varepsilon }_{c}})\,
       ({{\varepsilon }_{mw}}-{{\varepsilon }_{ab}})\,
       ({{\varepsilon }_{nrw}}-{{\varepsilon }_{abc}})}\,+
\]
\[
-\left( \frac{1}{[w]} \right) \sum_{abcmnr}\sum_{L_{1}L_{3}}
    \frac{\delta _{\kappa }(c,w)\,\delta _{\kappa }(m,r)\,
       {\left( -1 \right) }^{-a + b + L_{3} - n + w}\,X_{L_{1}}(mnab)\,
       Z_{L_{1}}(banr)\,Z_{L_{3}}(rwcm)\,\langle c||z||v\rangle}{
       ({{\varepsilon }_w}-{{\varepsilon }_{c}})\,
       ({{\varepsilon }_{mn}}-{{\varepsilon }_{ab}})\,
       ({{\varepsilon }_{nrw}}-{{\varepsilon }_{abc}})\,[L_{1}]\,[m]}\,+
\]
\[
\,\frac{1}{{\sqrt{[w]}}} \sum_{abcdmn}\sum_{L_{1}}
    \frac{\delta _{\kappa }(a,c)\,\delta _{\kappa }(d,w)\,
       {\left( -1 \right) }^{a + b + m - n}\,X_{L_{1}}(mnab)\,Z_{0}(awcd)\,
       Z_{L_{1}}(bcnm)\,\langle d||z||v\rangle}{({{\varepsilon }_w}-
        {{\varepsilon }_{d}})\,
       ({{\varepsilon }_{mn}}-{{\varepsilon }_{ab}})\,
       ({{\varepsilon }_{mnw}}-{{\varepsilon }_{bcd}})\,{\sqrt{[a]}}\,[L_{1}]}
\,+
\]
\[
\frac{1}{[w]}\sum_{abcdmn}\sum_{L_{1}L_{2}}
    \frac{\delta _{\kappa }(d,w)\,
       {\left( -1 \right) }^{-a + b + c + L_{2} - m - n - w}\,
       Z_{L_{1}}(bcmn)\,Z_{L_{1}}(mwba)\,Z_{L_{2}}(nacd)\,
       \langle d||z||v\rangle}{({{\varepsilon }_w}-{{\varepsilon }_{d}})\,
       ({{\varepsilon }_{mw}}-{{\varepsilon }_{ab}})\,
       ({{\varepsilon }_{mnw}}-{{\varepsilon }_{bcd}})\,{[L_{2}]}^2}\, +
\]
\[
\overline{h.c.s.}
\]

\newpage

\subsection{$Z_{0\times 3}\left( D_v[T_c] \right)$}
\[
Z_{0\times 3}\left( D_v[T_c] \right)=
\]


\[
\frac{1}{[K]\,[v]}\sum_{abcdmn}\sum_{L_{1}}
    \frac{\delta _{\kappa }(a,v)\,
       {\left( -1 \right) }^{b - c + d + K - m - n - v}\,X_{L_{1}}(ancd)\,
       Z_{K}(mwba)\,Z_{L_{1}}(dcnv)\,\langle b||z||m\rangle}{(
        {{\varepsilon }_{mw}}-{{\varepsilon }_{ab}})\,
       ({{\varepsilon }_{mw}}-{{\varepsilon }_{bv}})\,
       ({{\varepsilon }_{mnw}}-{{\varepsilon }_{bcd}})\,[L_{1}]}\,+
\]
\[
\frac{1}{[K]}\sum_{abcdmn}\sum_{L_{1}L_{2}}
    \frac{{\left( -1 \right) }^{b - n - v - w}\,
       \left\{ \begin{array}[c]{ccc}K & v &w  \\  d & L_{1} & L_{2}
        \end{array} \right\}\,\left\{ \begin{array}[c]{ccc}L_{2} & K &
        L_{1}  \\  a & c & m\end{array} \right\}\,X_{L_{1}}(awcd)\,
       Z_{K}(mnab)\,Z_{L_{2}}(cdmv)\,\langle b||z||n\rangle}{(
        {{\varepsilon }_{mn}}-{{\varepsilon }_{ab}})\,
       ({{\varepsilon }_{nw}}-{{\varepsilon }_{bv}})\,
       ({{\varepsilon }_{mnw}}-{{\varepsilon }_{bcd}})}\,+
\]
\[
\sum_{abcdmn}\sum_{L_{1}L_{2}L_{3}}
   \frac{{\left( -1 \right) }^{-b + K + L_{3} - v}\,
      \left\{ \begin{array}[c]{ccc}L_{2} & L_{3} &L_{1}  \\  a & c & m
       \end{array} \right\}\,\left\{ \begin{array}[c]{ccc}L_{3} & v &n
         \\  d & L_{1} & L_{2}\end{array} \right\}\,
      \left\{ \begin{array}[c]{ccc}v & L_{3} &n  \\  b & K & w
       \end{array} \right\}\,X_{L_{1}}(ancd)\,Z_{L_{2}}(cdmv)\,
      Z_{L_{3}}(mwab)\,\langle b||z||n\rangle}{({{\varepsilon }_{mw}}-
       {{\varepsilon }_{ab}})\,
      ({{\varepsilon }_{nw}}-{{\varepsilon }_{bv}})\,
      ({{\varepsilon }_{mnw}}-{{\varepsilon }_{bcd}})}\,+
\]
\[
-\left( \frac{1}{[K]} \right) \sum_{abcmnr}\sum_{L_{1}}
    \frac{{\left( -1 \right) }^{a + b + c + L_{1} - m - n - r}\,
       \left\{ \begin{array}[c]{ccc}w & L_{1} &m  \\  a & K & v
        \end{array} \right\}\,Z_{K}(mnab)\,Z_{L_{1}}(carv)\,Z_{L_{1}}(rwcm)\,
       \langle b||z||n\rangle}{({{\varepsilon }_{mn}}-{{\varepsilon }_{ab}}
        )\,({{\varepsilon }_{nw}}-{{\varepsilon }_{bv}})\,
       ({{\varepsilon }_{nrw}}-{{\varepsilon }_{abc}})\,[L_{1}]}\,+
\]
\[
\sum_{abcmnr}\sum_{L_{1}L_{2}} \frac{{\left( -1 \right) }^
       {-a - b + c + K + L_{1} + L_{2} - m - n + r}\,
      \left\{ \begin{array}[c]{ccc}r & L_{2} &v  \\  a & L_{1} & m
       \end{array} \right\}\,\left\{ \begin{array}[c]{ccc}v & L_{2} &r
         \\  b & K & w\end{array} \right\}\,Z_{L_{1}}(canv)\,
      Z_{L_{1}}(nrcm)\,Z_{L_{2}}(mwab)\,\langle b||z||r\rangle}{
      ({{\varepsilon }_{mw}}-{{\varepsilon }_{ab}})\,
      ({{\varepsilon }_{rw}}-{{\varepsilon }_{bv}})\,
      ({{\varepsilon }_{nrw}}-{{\varepsilon }_{abc}})\,[L_{1}]}\,+
\]
\[
-\sum_{abcmnr}\sum_{L_{1}L_{3}}
    \frac{{\left( -1 \right) }^{a + K + L_{1} - n - v - w}\,
       \left\{ \begin{array}[c]{ccc}K & w &v  \\  c & L_{1} & L_{3}
        \end{array} \right\}\,\left\{ \begin{array}[c]{ccc}L_{3} & K &
        L_{1}  \\  b & m & r\end{array} \right\}\,Z_{L_{1}}(acnv)\,
       Z_{L_{1}}(nmab)\,Z_{L_{3}}(wrcm)\,\langle b||z||r\rangle}{
       ({{\varepsilon }_{mn}}-{{\varepsilon }_{ab}})\,
       ({{\varepsilon }_{rw}}-{{\varepsilon }_{bv}})\,
       ({{\varepsilon }_{nrw}}-{{\varepsilon }_{abc}})\,[L_{1}]}\,+
\]
\[
-\sum_{abcmnr}\sum_{L_{1}L_{2}L_{3}}
    \frac{{\left( -1 \right) }^{-c + K + L_{3} - v}\,
       \left\{ \begin{array}[c]{ccc}L_{3} & L_{2} &L_{1}  \\  a & m & r
        \end{array} \right\}\,\left\{ \begin{array}[c]{ccc}L_{3} & v &n
          \\  b & L_{1} & L_{2}\end{array} \right\}\,
       \left\{ \begin{array}[c]{ccc}v & L_{3} &n  \\  c & K & w
        \end{array} \right\}\,X_{L_{1}}(mnab)\,Z_{L_{2}}(bavr)\,
       Z_{L_{3}}(rwmc)\,\langle c||z||n\rangle}{({{\varepsilon }_{mn}}-
        {{\varepsilon }_{ab}})\,
       ({{\varepsilon }_{nw}}-{{\varepsilon }_{cv}})\,
       ({{\varepsilon }_{nrw}}-{{\varepsilon }_{abc}})}\,+
\]
\[
-\left( \frac{1}{[K]} \right) \sum_{abcmnr}\sum_{L_{1}L_{2}}
    \frac{{\left( -1 \right) }^{c - r - v - w}\,
       \left\{ \begin{array}[c]{ccc}K & L_{2} &L_{1}  \\  a & m & n
        \end{array} \right\}\,\left\{ \begin{array}[c]{ccc}K & v &w  \\  b
         & L_{1} & L_{2}\end{array} \right\}\,X_{L_{1}}(mwab)\,
       Z_{K}(nrmc)\,Z_{L_{2}}(abnv)\,\langle c||z||r\rangle}{(
        {{\varepsilon }_{mw}}-{{\varepsilon }_{ab}})\,
       ({{\varepsilon }_{rw}}-{{\varepsilon }_{cv}})\,
       ({{\varepsilon }_{nrw}}-{{\varepsilon }_{abc}})}\,+
\]
\[
-\left( \frac{1}{[K]\,[v]} \right)
   \sum_{abcmnr}\sum_{L_{1}} \frac{\delta _{\kappa }(m,v)\,
       {\left( -1 \right) }^{-a + b + c + K - n - r - v}\,X_{L_{1}}(mnab)\,
       Z_{K}(rwcm)\,Z_{L_{1}}(banv)\,\langle c||z||r\rangle}{(
        {{\varepsilon }_{mn}}-{{\varepsilon }_{ab}})\,
       ({{\varepsilon }_{rw}}-{{\varepsilon }_{cv}})\,
       ({{\varepsilon }_{nrw}}-{{\varepsilon }_{abc}})\,[L_{1}]}\,+
\]
\[
\sum_{abcdmn}\sum_{L_{1}L_{2}} \frac{{\left( -1 \right) }^
       {c + K + L_{1} - n - v - w}\,
      \left\{ \begin{array}[c]{ccc}K & L_{2} &L_{1}  \\  a & d & m
       \end{array} \right\}\,\left\{ \begin{array}[c]{ccc}K & v &w  \\  b
        & L_{2} & L_{1}\end{array} \right\}\,Z_{L_{1}}(cbnv)\,
      Z_{L_{1}}(nacd)\,Z_{L_{2}}(mwab)\,\langle d||z||m\rangle}{
      ({{\varepsilon }_{mw}}-{{\varepsilon }_{ab}})\,
      ({{\varepsilon }_{mw}}-{{\varepsilon }_{dv}})\,
      ({{\varepsilon }_{mnw}}-{{\varepsilon }_{bcd}})\,[L_{1}]}\,+
\]
\[
-\sum_{abcdmn}\sum_{L_{1}L_{2}}
    \frac{{\left( -1 \right) }^{-a + b - c - d + K + L_{1} + L_{2} - m + n}\,
       \left\{ \begin{array}[c]{ccc}v & L_{1} &n  \\  d & K & w
        \end{array} \right\}\,\left\{ \begin{array}[c]{ccc}v & n &L_{1}
          \\  a & c & L_{2}\end{array} \right\}\,Z_{L_{1}}(awcd)\,
       Z_{L_{2}}(bcmv)\,Z_{L_{2}}(mnba)\,\langle d||z||n\rangle}{
       ({{\varepsilon }_{mn}}-{{\varepsilon }_{ab}})\,
       ({{\varepsilon }_{nw}}-{{\varepsilon }_{dv}})\,
       ({{\varepsilon }_{mnw}}-{{\varepsilon }_{bcd}})\,[L_{2}]}\,+
\]
\[
-\left( \frac{1}{[K]} \right) \sum_{abcdmn}\sum_{L_{2}}
    \frac{{\left( -1 \right) }^{-a + b - c + d + L_{2} - m - n}\,
       \left\{ \begin{array}[c]{ccc}v & w &K  \\  a & c & L_{2}
        \end{array} \right\}\,Z_{K}(ancd)\,Z_{L_{2}}(bcmv)\,Z_{L_{2}}(mwba)\,
       \langle d||z||n\rangle}{({{\varepsilon }_{mw}}-{{\varepsilon }_{ab}}
        )\,({{\varepsilon }_{nw}}-{{\varepsilon }_{dv}})\,
       ({{\varepsilon }_{mnw}}-{{\varepsilon }_{bcd}})\,[L_{2}]}\,. +
\]
\[
\overline{h.c.s.}
\]

\newpage

\subsection{$Z_{0\times 3}\left( D_v[T_v^h] \right)$}
\[
Z_{0\times 3}\left( D_v[T_v^h] \right) =
\]

\[
-\left( \frac{1}{[K]} \right) \sum_{abcmnr}\sum_{L_{1}}
    \frac{{\left( -1 \right) }^{a + b + c + L_{1} - m - n - r}\,
       \left\{ \begin{array}[c]{ccc}w & L_{1} &m  \\  a & K & v
        \end{array} \right\}\,Z_{K}(mnab)\,Z_{L_{1}}(cwrm)\,Z_{L_{1}}(racv)\,
       \langle b||z||n\rangle}{({{\varepsilon }_{mn}}-{{\varepsilon }_{ab}}
        )\,({{\varepsilon }_{nw}}-{{\varepsilon }_{bv}})\,
       ({{\varepsilon }_{mnr}}-{{\varepsilon }_{bcv}})\,[L_{1}]}\,+
\]
\[
\frac{1}{[K]}\sum_{abcmnr}\sum_{L_{1}L_{3}}
    \frac{\left\{ \begin{array}[c]{ccc}L_{3} & L_{1} &K  \\  a & c & m
        \end{array} \right\}\,\left\{ \begin{array}[c]{ccc}L_{3} & r &n
          \\  b & L_{1} & K\end{array} \right\}\,X_{L_{1}}(mnab)\,
       Z_{K}(awcv)\,Z_{L_{3}}(crmn)\,\langle b||z||r\rangle}{(
        {{\varepsilon }_{mn}}-{{\varepsilon }_{ab}})\,
       ({{\varepsilon }_{rw}}-{{\varepsilon }_{bv}})\,
       ({{\varepsilon }_{mnw}}-{{\varepsilon }_{bcv}})}\,+
\]
\[
\sum_{abcmnr}\sum_{L_{1}L_{2}L_{3}}
   \frac{{\left( -1 \right) }^{K + L_{2} - r + w}\,
      \left\{ \begin{array}[c]{ccc}L_{2} & w &b  \\  K & r & v
       \end{array} \right\}\,\left\{ \begin{array}[c]{ccc}L_{3} & L_{1} &
       L_{2}  \\  a & c & m\end{array} \right\}\,
      \left\{ \begin{array}[c]{ccc}L_{3} & w &n  \\  b & L_{1} & L_{2}
       \end{array} \right\}\,X_{L_{1}}(mnab)\,Z_{L_{2}}(arcv)\,
      Z_{L_{3}}(cwmn)\,\langle b||z||r\rangle}{({{\varepsilon }_{mn}}-
       {{\varepsilon }_{ab}})\,
      ({{\varepsilon }_{rw}}-{{\varepsilon }_{bv}})\,
      ({{\varepsilon }_{mnr}}-{{\varepsilon }_{bcv}})}\,+
\]
\[
\sum_{abcmnr}\sum_{L_{1}L_{2}} \frac{{\left( -1 \right) }^
       {-a - b + c + K + L_{1} + L_{2} - m - n + r}\,
      \left\{ \begin{array}[c]{ccc}r & L_{2} &v  \\  a & L_{1} & m
       \end{array} \right\}\,\left\{ \begin{array}[c]{ccc}v & L_{2} &r
         \\  b & K & w\end{array} \right\}\,Z_{L_{1}}(crnm)\,
      Z_{L_{1}}(nacv)\,Z_{L_{2}}(mwab)\,\langle b||z||r\rangle}{
      ({{\varepsilon }_{mw}}-{{\varepsilon }_{ab}})\,
      ({{\varepsilon }_{rw}}-{{\varepsilon }_{bv}})\,
      ({{\varepsilon }_{mnw}}-{{\varepsilon }_{bcv}})\,[L_{1}]}\,+
\]
\[
-\sum_{abcmnr}\sum_{L_{1}L_{2}}
    \frac{{\left( -1 \right) }^{b + K + L_{2} - m + v + w}\,
       \left\{ \begin{array}[c]{ccc}K & L_{2} &L_{1}  \\  a & c & n
        \end{array} \right\}\,\left\{ \begin{array}[c]{ccc}r & w &L_{2}
          \\  K & L_{1} & v\end{array} \right\}\,Z_{L_{1}}(arcv)\,
       Z_{L_{2}}(bwmr)\,Z_{L_{2}}(mnba)\,\langle c||z||n\rangle}{
       ({{\varepsilon }_{mn}}-{{\varepsilon }_{ab}})\,
       ({{\varepsilon }_{nw}}-{{\varepsilon }_{cv}})\,
       ({{\varepsilon }_{mnr}}-{{\varepsilon }_{bcv}})\,[L_{2}]}\,+
\]
\[
\sum_{abcmnr}\sum_{L_{1}L_{2}} \frac{{\left( -1 \right) }^
       {-a + b - c + K + L_{1} + L_{2} - m + n - r}\,
      \left\{ \begin{array}[c]{ccc}v & L_{2} &n  \\  c & K & w
       \end{array} \right\}\,\left\{ \begin{array}[c]{ccc}w & L_{2} &c
         \\  a & L_{1} & m\end{array} \right\}\,Z_{L_{1}}(bwrm)\,
      Z_{L_{1}}(rabc)\,Z_{L_{2}}(mnav)\,\langle c||z||n\rangle}{
      ({{\varepsilon }_{mn}}-{{\varepsilon }_{av}})\,
      ({{\varepsilon }_{nw}}-{{\varepsilon }_{cv}})\,
      ({{\varepsilon }_{mnr}}-{{\varepsilon }_{bcv}})\,[L_{1}]}\,+
\]
\[
\sum_{abcmnr}\sum_{L_{1}L_{2}L_{3}}
   \frac{{\left( -1 \right) }^{-c + K + L_{2} + v}\,
      \left\{ \begin{array}[c]{ccc}L_{3} & L_{1} &L_{2}  \\  a & b & m
       \end{array} \right\}\,\left\{ \begin{array}[c]{ccc}r & n &L_{3}
         \\  L_{1} & L_{2} & v\end{array} \right\}\,
      \left\{ \begin{array}[c]{ccc}v & L_{2} &r  \\  c & K & w
       \end{array} \right\}\,X_{L_{1}}(mnav)\,Z_{L_{2}}(awbc)\,
      Z_{L_{3}}(brmn)\,\langle c||z||r\rangle}{({{\varepsilon }_{mn}}-
       {{\varepsilon }_{av}})\,
      ({{\varepsilon }_{rw}}-{{\varepsilon }_{cv}})\,
      ({{\varepsilon }_{mnw}}-{{\varepsilon }_{bcv}})}\,+
\]
\[
-\left( \frac{1}{[K]} \right) \sum_{abcmnr}\sum_{L_{1}}
    \frac{\delta _{\kappa }(a,r)\,{\left( -1 \right) }^{b + c + K + m - n}\,
       X_{L_{1}}(mnab)\,Z_{K}(awcv)\,Z_{L_{1}}(brnm)\,\langle c||z||r\rangle}
{({{\varepsilon }_{mn}}-{{\varepsilon }_{ab}})\,
       ({{\varepsilon }_{rw}}-{{\varepsilon }_{cv}})\,
       ({{\varepsilon }_{mnw}}-{{\varepsilon }_{bcv}})\,[a]\,[L_{1}]}\,+
\]
\[
-\left( \frac{1}{[K]} \right) \sum_{abcmnr}\sum_{L_{1}L_{3}}
    \frac{{\left( -1 \right) }^{c - r - v - w}\,
       \left\{ \begin{array}[c]{ccc}L_{3} & L_{1} &K  \\  a & b & m
        \end{array} \right\}\,\left\{ \begin{array}[c]{ccc}w & n &L_{3}
          \\  L_{1} & K & v\end{array} \right\}\,X_{L_{1}}(mnav)\,
       Z_{K}(arbc)\,Z_{L_{3}}(bwmn)\,\langle c||z||r\rangle}{(
        {{\varepsilon }_{mn}}-{{\varepsilon }_{av}})\,
       ({{\varepsilon }_{rw}}-{{\varepsilon }_{cv}})\,
       ({{\varepsilon }_{mnr}}-{{\varepsilon }_{bcv}})}\,+
\]
\[
-\left( \frac{1}{[K]\,[w]} \right)
   \sum_{abcmnr}\sum_{L_{1}} \frac{\delta _{\kappa }(a,w)\,
       {\left( -1 \right) }^{b + c + K + m - n - r + w}\,X_{L_{1}}(mnab)\,
       Z_{K}(arvc)\,Z_{L_{1}}(bwnm)\,\langle c||z||r\rangle}{(
        {{\varepsilon }_{mn}}-{{\varepsilon }_{ab}})\,
       ({{\varepsilon }_{rw}}-{{\varepsilon }_{cv}})\,
       ({{\varepsilon }_{mnr}}-{{\varepsilon }_{bcv}})\,[L_{1}]}\,+
\]
\[
\sum_{abcmnr}\sum_{L_{1}L_{3}} \frac{{\left( -1 \right) }^
       {b + K + L_{1} - m + v - w}\,
      \left\{ \begin{array}[c]{ccc}K & L_{1} &L_{3}  \\  a & v & w
       \end{array} \right\}\,\left\{ \begin{array}[c]{ccc}L_{1} & n &r
         \\  c & K & L_{3}\end{array} \right\}\,Z_{L_{1}}(brmn)\,
      Z_{L_{1}}(mwba)\,Z_{L_{3}}(nacv)\,\langle c||z||r\rangle}{
      ({{\varepsilon }_{mw}}-{{\varepsilon }_{ab}})\,
      ({{\varepsilon }_{rw}}-{{\varepsilon }_{cv}})\,
      ({{\varepsilon }_{mnw}}-{{\varepsilon }_{bcv}})\,[L_{1}]}\,+
\]
\[
\frac{1}{[K]}\sum_{abcmnr}\sum_{L_{1}}
    \frac{{\left( -1 \right) }^{-a + b - c + L_{1} + m - n - r}\,
       \left\{ \begin{array}[c]{ccc}r & m &L_{1}  \\  a & c & K
        \end{array} \right\}\,Z_{K}(mwav)\,Z_{L_{1}}(brnm)\,Z_{L_{1}}(nabc)\,
       \langle c||z||r\rangle}{({{\varepsilon }_{mw}}-{{\varepsilon }_{av}}
        )\,({{\varepsilon }_{rw}}-{{\varepsilon }_{cv}})\,
       ({{\varepsilon }_{mnw}}-{{\varepsilon }_{bcv}})\,[L_{1}]}\,+
\]
\[
-\sum_{abcdmn}\sum_{L_{1}L_{3}}
    \frac{{\left( -1 \right) }^{c + K + L_{1} - n + v - w}\,
       \left\{ \begin{array}[c]{ccc}K & L_{3} &L_{1}  \\  a & v & w
        \end{array} \right\}\,\left\{ \begin{array}[c]{ccc}K & L_{3} &
        L_{1}  \\  b & d & m\end{array} \right\}\,Z_{L_{1}}(anvc)\,
       Z_{L_{1}}(bcdn)\,Z_{L_{3}}(mwba)\,\langle d||z||m\rangle}{
       ({{\varepsilon }_{mw}}-{{\varepsilon }_{ab}})\,
       ({{\varepsilon }_{mw}}-{{\varepsilon }_{dv}})\,
       ({{\varepsilon }_{mnw}}-{{\varepsilon }_{bcv}})\,[L_{1}]}\,+
\]
\[
\frac{1}{[K]}\sum_{abcdmn}\sum_{L_{1}}
    \frac{\delta _{\kappa }(a,d)\,{\left( -1 \right) }^{-b + c + K - m - n}\,
       X_{L_{1}}(anbc)\,Z_{K}(mwav)\,Z_{L_{1}}(bcdn)\,\langle d||z||m\rangle}
{({{\varepsilon }_{mw}}-{{\varepsilon }_{av}})\,
       ({{\varepsilon }_{mw}}-{{\varepsilon }_{dv}})\,
       ({{\varepsilon }_{mnw}}-{{\varepsilon }_{bcv}})\,[a]\,[L_{1}]}\,+
\]
\[
-\left( \frac{1}{[K]} \right) \sum_{abcdmn}\sum_{L_{2}}
    \frac{{\left( -1 \right) }^{-a + b - c - d + L_{2} - m + n}\,
       \left\{ \begin{array}[c]{ccc}d & n &K  \\  a & c & L_{2}
        \end{array} \right\}\,Z_{K}(awcv)\,Z_{L_{2}}(bcmd)\,Z_{L_{2}}(mnba)\,
       \langle d||z||n\rangle}{({{\varepsilon }_{mn}}-{{\varepsilon }_{ab}}
        )\,({{\varepsilon }_{nw}}-{{\varepsilon }_{dv}})\,
       ({{\varepsilon }_{mnw}}-{{\varepsilon }_{bcv}})\,[L_{2}]}\,+
\]
\[
\sum_{abcdmn}\sum_{L_{1}L_{2}L_{3}}
   \frac{{\left( -1 \right) }^{K + L_{3} + n + w}\,
      \left\{ \begin{array}[c]{ccc}L_{2} & L_{3} &L_{1}  \\  a & b & m
       \end{array} \right\}\,\left\{ \begin{array}[c]{ccc}L_{3} & d &w
         \\  c & L_{1} & L_{2}\end{array} \right\}\,
      \left\{ \begin{array}[c]{ccc}v & L_{3} &n  \\  d & K & w
       \end{array} \right\}\,X_{L_{1}}(awbc)\,Z_{L_{2}}(bcmd)\,
      Z_{L_{3}}(nmva)\,\langle d||z||n\rangle}{({{\varepsilon }_{mn}}-
       {{\varepsilon }_{av}})\,
      ({{\varepsilon }_{nw}}-{{\varepsilon }_{dv}})\,
      ({{\varepsilon }_{mnw}}-{{\varepsilon }_{bcv}})}\,+
\]
\[
-\sum_{abcdmn}\sum_{L_{1}L_{2}}
    \frac{{\left( -1 \right) }^{-a + b - c - d + K + L_{1} + L_{2} - m + n}\,
       \left\{ \begin{array}[c]{ccc}d & w &L_{1}  \\  a & c & L_{2}
        \end{array} \right\}\,\left\{ \begin{array}[c]{ccc}v & L_{1} &n
          \\  d & K & w\end{array} \right\}\,Z_{L_{1}}(ancv)\,
       Z_{L_{2}}(bcmd)\,Z_{L_{2}}(mwba)\,\langle d||z||n\rangle}{
       ({{\varepsilon }_{mw}}-{{\varepsilon }_{ab}})\,
       ({{\varepsilon }_{nw}}-{{\varepsilon }_{dv}})\,
       ({{\varepsilon }_{mnw}}-{{\varepsilon }_{bcv}})\,[L_{2}]}\,+
\]
\[
-\left( \frac{1}{[K]} \right) \sum_{abcdmn}\sum_{L_{1}L_{2}}
    \frac{\left\{ \begin{array}[c]{ccc}K & L_{1} &L_{2}  \\  c & d & n
        \end{array} \right\}\,\left\{ \begin{array}[c]{ccc}L_{2} & K &
        L_{1}  \\  a & b & m\end{array} \right\}\,X_{L_{1}}(anbc)\,
       Z_{K}(mwav)\,Z_{L_{2}}(cbdm)\,\langle d||z||n\rangle}{(
        {{\varepsilon }_{mw}}-{{\varepsilon }_{av}})\,
       ({{\varepsilon }_{nw}}-{{\varepsilon }_{dv}})\,
       ({{\varepsilon }_{mnw}}-{{\varepsilon }_{bcv}})}\, +
\]
\[
\overline{h.c.s.}
\]

\newpage

\subsection{$Z_{0\times 3}\left( D_v[T_v^p] \right)$}
\[
Z_{0\times 3}\left( D_v[T_v^p] \right) =
\]

\[
\frac{1}{[K]\,[w]}\sum_{abmnrs}\sum_{L_{1}}
    \frac{\delta _{\kappa }(m,w)\,
       {\left( -1 \right) }^{a + b + K - n - r + s + w}\,X_{L_{1}}(rsbm)\,
       Z_{K}(mnva)\,Z_{L_{1}}(bwrs)\,\langle a||z||n\rangle}{(
        {{\varepsilon }_{mn}}-{{\varepsilon }_{av}})\,
       ({{\varepsilon }_{nw}}-{{\varepsilon }_{av}})\,
       ({{\varepsilon }_{nrs}}-{{\varepsilon }_{abv}})\,[L_{1}]}\,+
\]
\[
\frac{1}{[K]}\sum_{abmnrs}\sum_{L_{1}}
    \frac{\delta _{\kappa }(m,s)\,{\left( -1 \right) }^{a + b + K - n + r}\,
       X_{L_{1}}(nrbm)\,Z_{K}(mwav)\,Z_{L_{1}}(bsnr)\,\langle a||z||s\rangle}
{({{\varepsilon }_{mw}}-{{\varepsilon }_{av}})\,
       ({{\varepsilon }_{sw}}-{{\varepsilon }_{av}})\,
       ({{\varepsilon }_{nrw}}-{{\varepsilon }_{abv}})\,[L_{1}]\,[m]}\,+
\]
\[
\sum_{abmnrs}\sum_{L_{1}L_{2}} \frac{{\left( -1 \right) }^
       {b + K + L_{1} - r + v + w}\,
      \left\{ \begin{array}[c]{ccc}L_{1} & K &L_{2}  \\  a & m & s
       \end{array} \right\}\,\left\{ \begin{array}[c]{ccc}n & w &L_{1}
         \\  K & L_{2} & v\end{array} \right\}\,Z_{L_{1}}(bwrn)\,
      Z_{L_{1}}(rsbm)\,Z_{L_{2}}(mnav)\,\langle a||z||s\rangle}{
      ({{\varepsilon }_{mn}}-{{\varepsilon }_{av}})\,
      ({{\varepsilon }_{sw}}-{{\varepsilon }_{av}})\,
      ({{\varepsilon }_{nrs}}-{{\varepsilon }_{abv}})\,[L_{1}]}\,+
\]
\[
\sum_{abmnrs}\sum_{L_{1}L_{2}} \frac{{\left( -1 \right) }^
       {b + K + L_{1} - r + v + w}\,
      \left\{ \begin{array}[c]{ccc}L_{1} & K &L_{2}  \\  a & n & s
       \end{array} \right\}\,\left\{ \begin{array}[c]{ccc}m & w &L_{1}
         \\  K & L_{2} & v\end{array} \right\}\,Z_{L_{1}}(bsrn)\,
      Z_{L_{1}}(rwbm)\,Z_{L_{2}}(nmav)\,\langle a||z||s\rangle}{
      ({{\varepsilon }_{mn}}-{{\varepsilon }_{av}})\,
      ({{\varepsilon }_{sw}}-{{\varepsilon }_{av}})\,
      ({{\varepsilon }_{nrw}}-{{\varepsilon }_{abv}})\,[L_{1}]}\,+
\]
\[
\frac{1}{[K]}\sum_{abmnrs}\sum_{L_{1}L_{2}}
    \frac{{\left( -1 \right) }^{b - n - v - w}\,
       \left\{ \begin{array}[c]{ccc}L_{1} & L_{2} &K  \\  a & m & r
        \end{array} \right\}\,\left\{ \begin{array}[c]{ccc}w & v &K  \\
        L_{1} & L_{2} & s\end{array} \right\}\,X_{L_{1}}(rsmv)\,
       Z_{K}(mnab)\,Z_{L_{2}}(awrs)\,\langle b||z||n\rangle}{(
        {{\varepsilon }_{mn}}-{{\varepsilon }_{ab}})\,
       ({{\varepsilon }_{nw}}-{{\varepsilon }_{bv}})\,
       ({{\varepsilon }_{nrs}}-{{\varepsilon }_{abv}})}\,+
\]
\[
-\sum_{abmnrs}\sum_{L_{1}L_{2}L_{3}}
    \frac{{\left( -1 \right) }^{K + L_{3} - n + w}\,
       \left\{ \begin{array}[c]{ccc}L_{1} & L_{2} &L_{3}  \\  a & m & s
        \end{array} \right\}\,\left\{ \begin{array}[c]{ccc}L_{2} & w &r
          \\  b & L_{1} & L_{3}\end{array} \right\}\,
       \left\{ \begin{array}[c]{ccc}L_{3} & w &b  \\  K & n & v
        \end{array} \right\}\,X_{L_{1}}(rsbm)\,Z_{L_{2}}(awsr)\,
       Z_{L_{3}}(mnav)\,\langle b||z||n\rangle}{({{\varepsilon }_{mn}}-
        {{\varepsilon }_{av}})\,
       ({{\varepsilon }_{nw}}-{{\varepsilon }_{bv}})\,
       ({{\varepsilon }_{nrs}}-{{\varepsilon }_{abv}})}\,+
\]
\[
\sum_{abmnrs}\sum_{L_{1}L_{2}L_{3}}
   \frac{{\left( -1 \right) }^{-b + K + L_{3} - v}\,
      \left\{ \begin{array}[c]{ccc}L_{1} & L_{2} &L_{3}  \\  a & m & n
       \end{array} \right\}\,\left\{ \begin{array}[c]{ccc}s & v &L_{3}
         \\  L_{1} & L_{2} & r\end{array} \right\}\,
      \left\{ \begin{array}[c]{ccc}v & L_{3} &s  \\  b & K & w
       \end{array} \right\}\,X_{L_{1}}(nrmv)\,Z_{L_{2}}(asnr)\,
      Z_{L_{3}}(mwab)\,\langle b||z||s\rangle}{({{\varepsilon }_{mw}}-
       {{\varepsilon }_{ab}})\,
      ({{\varepsilon }_{sw}}-{{\varepsilon }_{bv}})\,
      ({{\varepsilon }_{nrw}}-{{\varepsilon }_{abv}})}\,+
\]
\[
-\left( \frac{1}{[K]} \right) \sum_{abmnrs}\sum_{L_{1}L_{2}}
    \frac{\left\{ \begin{array}[c]{ccc}L_{1} & L_{2} &K  \\  a & m & r
        \end{array} \right\}\,\left\{ \begin{array}[c]{ccc}L_{2} & s &n
          \\  b & L_{1} & K\end{array} \right\}\,X_{L_{1}}(nrbm)\,
       Z_{K}(mwav)\,Z_{L_{2}}(asrn)\,\langle b||z||s\rangle}{(
        {{\varepsilon }_{mw}}-{{\varepsilon }_{av}})\,
       ({{\varepsilon }_{sw}}-{{\varepsilon }_{bv}})\,
       ({{\varepsilon }_{nrw}}-{{\varepsilon }_{abv}})}\,+
\]
\[
-\left( \frac{1}{[K]} \right) \sum_{abmnrs}\sum_{L_{1}}
    \frac{{\left( -1 \right) }^{a + b + L_{1} + m - n + r - s}\,
       \left\{ \begin{array}[c]{ccc}K & w &v  \\  L_{1} & m & r
        \end{array} \right\}\,Z_{K}(rsmb)\,Z_{L_{1}}(awnr)\,Z_{L_{1}}(nmav)\,
       \langle b||z||s\rangle}{({{\varepsilon }_{mn}}-{{\varepsilon }_{av}}
        )\,({{\varepsilon }_{sw}}-{{\varepsilon }_{bv}})\,
       ({{\varepsilon }_{nrs}}-{{\varepsilon }_{abv}})\,[L_{1}]}\,+
\]
\[
\sum_{abmnrs}\sum_{L_{1}L_{3}} \frac{{\left( -1 \right) }^
       {a + b + K + L_{1} + L_{3} - m - n + r - s}\,
      \left\{ \begin{array}[c]{ccc}L_{3} & w &b  \\  K & s & v
       \end{array} \right\}\,\left\{ \begin{array}[c]{ccc}r & L_{3} &m
         \\  b & L_{1} & w\end{array} \right\}\,Z_{L_{1}}(awnr)\,
      Z_{L_{1}}(nmab)\,Z_{L_{3}}(rsmv)\,\langle b||z||s\rangle}{
      ({{\varepsilon }_{mn}}-{{\varepsilon }_{ab}})\,
      ({{\varepsilon }_{sw}}-{{\varepsilon }_{bv}})\,
      ({{\varepsilon }_{nrs}}-{{\varepsilon }_{abv}})\,[L_{1}]}\,+
\]
\[
\sum_{abmnrs}\sum_{L_{1}L_{3}} \frac{{\left( -1 \right) }^
       {a + L_{3} - n + v + w}\,
      \left\{ \begin{array}[c]{ccc}L_{1} & K &L_{3}  \\  b & r & s
       \end{array} \right\}\,\left\{ \begin{array}[c]{ccc}m & w &L_{3}
         \\  K & L_{1} & v\end{array} \right\}\,Z_{L_{1}}(asnr)\,
      Z_{L_{1}}(nmav)\,Z_{L_{3}}(rwbm)\,\langle b||z||s\rangle}{
      ({{\varepsilon }_{mn}}-{{\varepsilon }_{av}})\,
      ({{\varepsilon }_{sw}}-{{\varepsilon }_{bv}})\,
      ({{\varepsilon }_{nrw}}-{{\varepsilon }_{abv}})\,[L_{1}]}\,+
\]
\[
-\left( \frac{1}{[K]} \right) \sum_{abmnrs}\sum_{L_{1}}
    \frac{{\left( -1 \right) }^{a + b + L_{1} - m - n + r + s}\,
       \left\{ \begin{array}[c]{ccc}r & K &m  \\  b & L_{1} & s
        \end{array} \right\}\,Z_{K}(rwmv)\,Z_{L_{1}}(asnr)\,Z_{L_{1}}(nmab)\,
       \langle b||z||s\rangle}{({{\varepsilon }_{mn}}-{{\varepsilon }_{ab}}
        )\,({{\varepsilon }_{sw}}-{{\varepsilon }_{bv}})\,
       ({{\varepsilon }_{nrw}}-{{\varepsilon }_{abv}})\,[L_{1}]}\,+
\]
\[
\sum_{abcmnr}\sum_{L_{1}L_{2}} \frac{{\left( -1 \right) }^
       {-a + b - c + K + L_{1} + L_{2} - m + n - r}\,
      \left\{ \begin{array}[c]{ccc}v & L_{2} &n  \\  c & K & w
       \end{array} \right\}\,\left\{ \begin{array}[c]{ccc}w & L_{2} &c
         \\  a & L_{1} & m\end{array} \right\}\,Z_{L_{1}}(abcr)\,
      Z_{L_{1}}(rwbm)\,Z_{L_{2}}(mnav)\,\langle c||z||n\rangle}{
      ({{\varepsilon }_{mn}}-{{\varepsilon }_{av}})\,
      ({{\varepsilon }_{nw}}-{{\varepsilon }_{cv}})\,
      ({{\varepsilon }_{nrw}}-{{\varepsilon }_{abv}})\,[L_{1}]}\,+
\]
\[
\frac{1}{[K]}\sum_{abcmnr}\sum_{L_{1}L_{2}}
    \frac{\left\{ \begin{array}[c]{ccc}K & L_{1} &L_{2}  \\  b & c & n
        \end{array} \right\}\,\left\{ \begin{array}[c]{ccc}K & L_{2} &
        L_{1}  \\  a & m & r\end{array} \right\}\,X_{L_{1}}(mnab)\,
       Z_{K}(rwmv)\,Z_{L_{2}}(abrc)\,\langle c||z||n\rangle}{(
        {{\varepsilon }_{mn}}-{{\varepsilon }_{ab}})\,
       ({{\varepsilon }_{nw}}-{{\varepsilon }_{cv}})\,
       ({{\varepsilon }_{nrw}}-{{\varepsilon }_{abv}})}\,+
\]
\[
\frac{1}{[K]}\sum_{abcmnr}\sum_{L_{1}}
    \frac{{\left( -1 \right) }^{-a + b - c + L_{1} + m - n - r}\,
       \left\{ \begin{array}[c]{ccc}r & m &L_{1}  \\  a & c & K
        \end{array} \right\}\,Z_{K}(mwav)\,Z_{L_{1}}(abcn)\,Z_{L_{1}}(nrbm)\,
       \langle c||z||r\rangle}{({{\varepsilon }_{mw}}-{{\varepsilon }_{av}}
        )\,({{\varepsilon }_{rw}}-{{\varepsilon }_{cv}})\,
       ({{\varepsilon }_{nrw}}-{{\varepsilon }_{abv}})\,[L_{1}]}\,+
\]
\[
-\sum_{abcmnr}\sum_{L_{1}L_{2}L_{3}}
    \frac{{\left( -1 \right) }^{K + L_{3} + r + w}\,
       \left\{ \begin{array}[c]{ccc}L_{3} & c &w  \\  b & L_{1} & L_{2}
        \end{array} \right\}\,\left\{ \begin{array}[c]{ccc}L_{3} & L_{2} &
        L_{1}  \\  a & m & n\end{array} \right\}\,
       \left\{ \begin{array}[c]{ccc}v & L_{3} &r  \\  c & K & w
        \end{array} \right\}\,X_{L_{1}}(mwab)\,Z_{L_{2}}(bacn)\,
       Z_{L_{3}}(rnvm)\,\langle c||z||r\rangle}{({{\varepsilon }_{mw}}-
        {{\varepsilon }_{ab}})\,
       ({{\varepsilon }_{rw}}-{{\varepsilon }_{cv}})\,
       ({{\varepsilon }_{nrw}}-{{\varepsilon }_{abv}})}\,+
\]
\[
-\sum_{abcmnr}\sum_{L_{1}L_{3}}
    \frac{{\left( -1 \right) }^{a + K + L_{1} - n + v + w}\,
       \left\{ \begin{array}[c]{ccc}K & L_{3} &L_{1}  \\  b & c & r
        \end{array} \right\}\,\left\{ \begin{array}[c]{ccc}m & w &L_{3}
          \\  K & L_{1} & v\end{array} \right\}\,Z_{L_{1}}(abnc)\,
       Z_{L_{1}}(nmav)\,Z_{L_{3}}(rwbm)\,\langle c||z||r\rangle}{
       ({{\varepsilon }_{mn}}-{{\varepsilon }_{av}})\,
       ({{\varepsilon }_{rw}}-{{\varepsilon }_{cv}})\,
       ({{\varepsilon }_{nrw}}-{{\varepsilon }_{abv}})\,[L_{1}]}\,+
\]
\[
-\left( \frac{1}{[K]} \right) \sum_{abcmnr}\sum_{L_{1}}
    \frac{\delta _{\kappa }(c,m)\,{\left( -1 \right) }^{-a + b + K - n - r}\,
       X_{L_{1}}(mnab)\,Z_{K}(rwmv)\,Z_{L_{1}}(abcn)\,\langle c||z||r\rangle}
{({{\varepsilon }_{mn}}-{{\varepsilon }_{ab}})\,
       ({{\varepsilon }_{rw}}-{{\varepsilon }_{cv}})\,
       ({{\varepsilon }_{nrw}}-{{\varepsilon }_{abv}})\,[c]\,[L_{1}]}\, +
\]
\[
\overline{h.c.s.}
\]

\newpage

\subsection{Normalization correction}
Finally, the angular reduction of normalization correction due to valence triple excitations is given by
\[
Z_\mathrm{norm}(T_v) =
- \frac{1}{2} \, \left( N_v^{(3)}(T_v) + N_w^{(3)}(T_v) \right) \, \langle w || z||v \rangle \,,
\]
with
\[
\frac{1}{2} N_v^{(3)}\left( T_v \right) =
\]

\[
\,\frac{1}{[v]} \sum_{abmnr}\sum_{L_{1}L_{2}L_{3}}
    \frac{\left\{ \begin{array}[c]{ccc}L_{2} & L_{3} &L_{1}  \\  a & m
         & n\end{array} \right\}\,
       \left\{ \begin{array}[c]{ccc}L_{2} & r &v  \\  b & L_{1} & L_{3}
        \end{array} \right\}\,X_{L_{1}}(mvab)\,X_{L_{2}}(nrmv)\,
       Z_{L_{3}}(abnr)}{({{\varepsilon }_{mv}}-{{\varepsilon }_{ab}})\,
       {({{\varepsilon }_{nr}}-{{\varepsilon }_{ab}})}^2}\,+
\]
\[
-\left( \frac{1}{[v]} \right) \sum_{abcmn}\sum_{L_{1}L_{2}L_{3}}
    \frac{\left\{ \begin{array}[c]{ccc}L_{2} & n &v  \\  c & L_{1} &
        L_{3}\end{array} \right\}\,
       \left\{ \begin{array}[c]{ccc}L_{3} & L_{2} &L_{1}  \\  a & b & m
        \end{array} \right\}\,X_{L_{1}}(avbc)\,X_{L_{2}}(mnav)\,
       Z_{L_{3}}(bcmn)}{({{\varepsilon }_{mn}}-{{\varepsilon }_{av}})\,
       {({{\varepsilon }_{mn}}-{{\varepsilon }_{bc}})}^2}\,+
\]
\[
-\,\frac{1}{{\sqrt{[v]}}} \sum_{abcmn}\sum_{L_{1}}
     \frac{\delta _{\kappa }(a,c)\,{\left( -1 \right) }^{a + b + m - n}\,
        X_{L_{1}}(mnab)\,Z_{0}(avcv)\,Z_{L_{1}}(bcnm)}{(
         {{\varepsilon }_{mn}}-{{\varepsilon }_{ab}})\,
        {({{\varepsilon }_{mn}}-{{\varepsilon }_{bc}})}^2\,{\sqrt{[a]}}\,
        [L_{1}]}\,+
\]
\[
\,\frac{1}{[v]} \sum_{abcmn}\sum_{L_{1}L_{2}}
    \frac{{\left( -1 \right) }^{a + b - c + L_{1} - m + n - v}\,
       Z_{L_{1}}(anvc)\,Z_{L_{1}}(mvba)\,Z_{L_{2}}(bcmn)}{{(
          {{\varepsilon }_{mn}}-{{\varepsilon }_{bc}})}^2\,
       ({{\varepsilon }_{mv}}-{{\varepsilon }_{ab}})\,{[L_{1}]}^2}\,+
\]
\[
\,\frac{1}{{\sqrt{[v]}}} \sum_{abmnr}\sum_{L_{1}}
    \frac{\delta _{\kappa }(a,m)\,{\left( -1 \right) }^{a + b - n + r}\,
       X_{L_{1}}(nrbm)\,Z_{0}(mvav)\,Z_{L_{1}}(banr)}{({{\varepsilon }_m}-
        {{\varepsilon }_{a}})\,
       {({{\varepsilon }_{nr}}-{{\varepsilon }_{ab}})}^2\,{\sqrt{[a]}}\,
       [L_{1}]}\,+
\]
\[
-\,\frac{1}{{\sqrt{[v]}}} \sum_{abcmn}\sum_{L_{1}}
     \frac{\delta _{\kappa }(a,m)\,{\left( -1 \right) }^{-a - b + c - n}\,
        X_{L_{1}}(anbc)\,Z_{0}(mvav)\,Z_{L_{1}}(bcmn)}{({{\varepsilon }_m}-
         {{\varepsilon }_{a}})\,
        {({{\varepsilon }_{mn}}-{{\varepsilon }_{bc}})}^2\,{\sqrt{[a]}}\,
        [L_{1}]}\,+
\]
\[
-\left( \frac{1}{[v]} \right) \sum_{abmnr}\sum_{L_{1}L_{2}}
    \frac{{\left( -1 \right) }^{a + b + L_{2} + m - n - r - v}\,
       Z_{L_{1}}(abnr)\,Z_{L_{1}}(nmav)\,Z_{L_{2}}(rvbm)}{(
        {{\varepsilon }_{mn}}-{{\varepsilon }_{av}})\,
       {({{\varepsilon }_{nr}}-{{\varepsilon }_{ab}})}^2\,{[L_{2}]}^2}\,+
\]
\[
\,\frac{1}{{\sqrt{[v]}}} \sum_{abmnr}\sum_{L_{1}}
    \frac{\delta _{\kappa }(m,r)\,{\left( -1 \right) }^{-a + b - m - n}\,
       X_{L_{1}}(mnab)\,Z_{0}(rvmv)\,Z_{L_{1}}(banr)}{({{\varepsilon }_{mn}}
        -{{\varepsilon }_{ab}})\,
       {({{\varepsilon }_{nr}}-{{\varepsilon }_{ab}})}^2\,[L_{1}]\,
       {\sqrt{[m]}}}\,.
\]

\end{document}